\documentstyle[multicol,aps,prl,psfig,epsf,epsfig]{revtex}

\begin{document}

\title{Response properties in a model for granular matter}
\author{A. Barrat $^{a}$ and V. Loreto $^{b}$}
\pagestyle{myheadings}
\address{$a)$Laboratoire de Physique Th{\'e}orique
\cite{umr}, B{\^a}timent 210, Universit{\'e}  
de Paris-Sud \\
91405 Orsay Cedex, France; e-mail: 
Alain.Barrat@th.u-psud.fr \\ 
$b)$ P.M.M.H. Ecole Sup\'erieure de Physique et Chimie Industrielles, \\
10, rue Vauquelin, 75231 Paris CEDEX 05 France;
e-mail: loreto@pmmh.espci.fr}

\maketitle
\begin{abstract} 
We investigate the response properties of granular media in the framework
of the so-called {\em Random Tetris Model}. We monitor,
for different driving procedures, several quantities: 
the evolution of the density and of the density profiles, the ageing 
properties through the two-times correlation functions and the two-times 
mean-square distance between the potential energies, the response function 
defined in terms  of the difference in the potential energies of two replica 
driven in two slightly different ways. We focus in particular on the role
played by the spatial inhomogeneities (structures) spontaneously emerging 
during the compaction process, the history of the sample and the driving 
procedure. It turns out that none of these ingredients can be neglected 
for the correct interpretation of the experimental or numerical data. 
We discuss the problem of the optimization of the compaction process and 
we comment on the validity of our results for the description of granular 
materials in a thermodynamic framework.

\end{abstract}
\smallskip


\section{introduction}

Granular media\cite{grain} are usually considered non-thermal systems
because their thermal energy is so negligibly small
with respect to other energy contributions
(e.g. potential energy)
that for all the practical purposes they
are virtually at zero temperature.
This feature draws a lot of consequences from the
point of view of the validity of thermodynamics
for such systems. One of the most important consequences
is that, unless perturbed in some way
(e.g. driving energy into the system), a granular system
cannot explore spontaneously its phase space but it remains
trapped in one of the numerous metastable configurations.
One has then to look at the dynamics of a granular system
always as a response to some perturbations and in general the response
will depend in a non-trivial way on the rheological properties of the
medium, on the boundaries, on the driving procedure and, last
but not the least, on the past history of the system.

In this paper we focus on the response properties of a class of lattice 
of  models, the so-called Random Tetris Model (RTM) \cite{RTM}, that, 
despite their apparent 
simplicity are able to reproduce many features of real granular materials:
slow-relaxations during compaction \cite{prltetris,RTM}, 
segregation\cite{segtet}, dilatancy properties \cite{dilatancy}, 
ageing \cite{nicodemi_coniglio}.

We shall be concerned in particular by the interplay between the 
response properties and the spatial structures 
that spontaneously
emerge as a consequence of the dynamics imposed to these systems.
Several examples in this direction have already been drawn: 
the phenomenon of structures formation that plays a crucial role 
in explaining the complex features of internal avalanching 
\cite{valanghe} or the coarsening phenomena that parallel the 
vibration-induced compaction process (to be discussed in a 
forthcoming paper \cite{baldassarri}).

We shall consider several procedures of vibro-compaction
and we shall study how the system responds to different procedures and 
how the emerging inhomogeneities affect the response properties of these 
systems. The vibration procedures are specified in terms of the
temporal function describing the evolution of the shaking amplitude.
The simplest case is the one where one keeps the shaking amplitude
indefinitely constant. More generally we shall consider complex
procedures corresponding to sequences of cooling and annealing
processes. 
In all the different cases we monitor several quantities.
On the one hand, we focus our attention on global quantities as
the global density, the response and the correlation functions.
On the other hand we monitor some local quantities that allow
us to investigate the large-scale structures spontaneously emerging
in these systems as a response to the imposed perturbation
(driving). The comparison between global and local quantities 
will be a valuable tool towards an understanding of how granular 
materials respond to perturbations and in this perspective 
of the importance  of spatial structures.  

In this framework we can address several questions. 
One of the first issues we can investigate is the 
importance of the history, i.e. the specific procedure undergone 
by the sample before we perform our measurements. It is interesting 
to ask where the history of the system is encoded and, for instance,
whether the global density represents a good parameter for the 
description of a static packing or one needs to specify other parameters.
The analysis of the correlation functions will allow us to 
investigate under which conditions the systems exhibit ageing
behaviour. On the other hand, with the knowledge of the effect 
of different kinds of perturbation, we shall address the problem of 
the optimization of the compaction procedures that can be stated as follows:
single out the best sequence of perturbations to impose to the system 
in order to maximize the density measured in a suitable part of 
the system.

Our results will also allow us to comment on the existence of transitions
in the response properties and on recently published results concerning
the violation of the fluctuation-dissipation theorem (FDT) for granular media 
\cite{nicodemi}.

The outline of the paper is as follows. In sect. II we recall the 
model definition, we define the different dynamical procedures and we 
introduce the quantities we shall look at for the analysis of the response 
properties. Sect. III and IV are devoted to the analysis of the 
response properties during a process of continuous shaking at constant 
amplitude and during cyclic procedures respectively.  
In sect. V we discuss globally all the results comparing different 
procedures and different histories and commenting on the consequences 
from the thermodynamic point of view. 
Finally in sect. VI we draw the conclusions.

\section{Model definition }

The essential ingredient of the RTM \cite{RTM} is the geometrical 
frustration
that for instance in granular packings is due to excluded volume 
effects arising from different shapes of the particles. 
This geometrical feature is captured in this 
class of lattice models where all the basic properties 
are brought by the particles and no assumptions are made on 
the environment (lattice). 
The interactions are not spatially quenched but are
determined in a self-consistent way by the local
arrangements of the particles. Despite the simplicity of
their definition, these systems present an highly complex phase-space and 
their dynamics generates automatically a very rich 
gallery of time-space correlations: time-scales, 
spatial structures, memory, etc.. Furthermore
they show a very interesting interplay between the dynamics 
and the time-space structures.
It is worth to notice how in this class of models
the origins of the randomness and of the frustration
coincide because both are given in terms of the particle
properties.

\begin{figure}[h]
\centerline{
       \psfig{figure=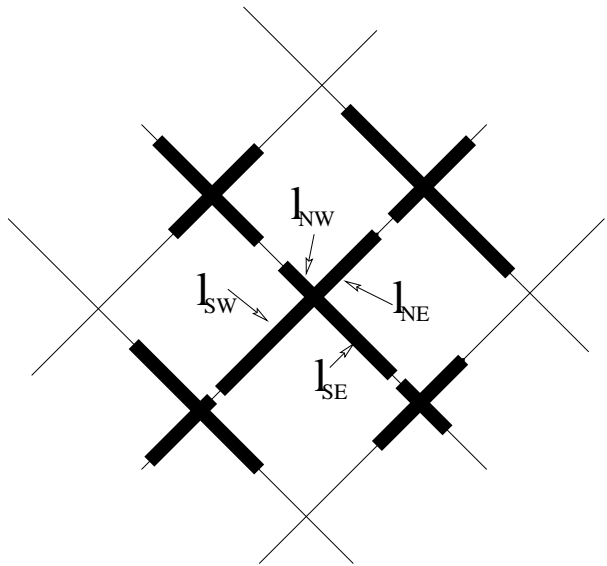,width=8cm,angle=0}
  \vspace{0.5cm}}
\caption{Sketch of a local arrangement of particles in the 
Random Tetris Model: each particles can be schematized in general
as a cross with  $4$ arms of different lengths,
denoted by $l_{NE}$, $l_{NW}$, $l_{SE}$, $l_{SW}$,
chosen in a random way.}            
\label{partrtm}
\end{figure}

Let us recall briefly the definition of the model, which includes,
like in the real computer game {\em Tetris}, a rich variety 
of shapes and sizes.
On a lattice each particle can be schematized in general
as a cross with  $4$ arms (in general the number of arms is equals 
to the coordination number of the lattice) of different lengths,
denoted by $l_{NE}$, $l_{NW}$, $l_{SE}$, $l_{SW}$, 
chosen in a random way. An example of particle configuration
on a tilted square lattice is shown in Fig.~\ref{partrtm}.

The static properties of the system are
then completely characterized giving the
random numbers defining the particles.
These numbers are given once for all and
it is clear how in this way one has a complete
freedom in the choice of the system and the models
used in \cite{prltetris,segtet} are particular cases
of the general model where one has chosen the
particle sizes and shapes in a deterministic way.

The interactions among the particles obey to the
general rule that one cannot have superpositions.
For instance one has to check that for two nearest
neighbours particles the sum of the arms oriented along
the bond connecting the two particles is smaller
than the bond length. It turns out that in this way
the interactions between the particles are not fixed
once for all but they depend on the complexity of the
spatial configuration.
We shall come back on this point later on in connection with
the interplay between the dynamics and the emergence of 
spatial structures.

The extreme generality of the model definition allows a large 
variety of choices for the particles.
Just to give the idea of how the system can be chosen
let us consider two different versions of the RTM
that we shall denote $A$ and $B$:

\begin{itemize}

\item[(A)] system with random elongated particles. This case represents
the direct generalization of the system considered in \cite{prltetris}.
Instead of considering two types of elongated particles with fixed sizes
we consider elongated particles whose size is chosen randomly according to
the expressions $l_{NE}=l_{SW}$, $l_{NW}=l_{SE}$,
$l_{NE}= \frac{3}{4} d \pm 0.1 d  \eta$, $l_{NW}=d - l_{NE}$
and equivalently for the other orientation. $\eta$ represents
a random variable uniformly distributed in $[0,1]$. 

\item[(B)] system with random particles with spherical symmetry. 
In this case we particles whose size is chosen randomly according to
the expressions $l_{NE}=l_{SW}$, $l_{NW}=l_{SE}$,
$l_{NE}= \frac{1}{2} d \pm 0.2 d \eta$, 
$l_{NW}= \frac{1}{2} d \pm 0.2 d \eta$.
 
\end{itemize}

\begin{figure}[h]
\centerline{
  \psfig{figure=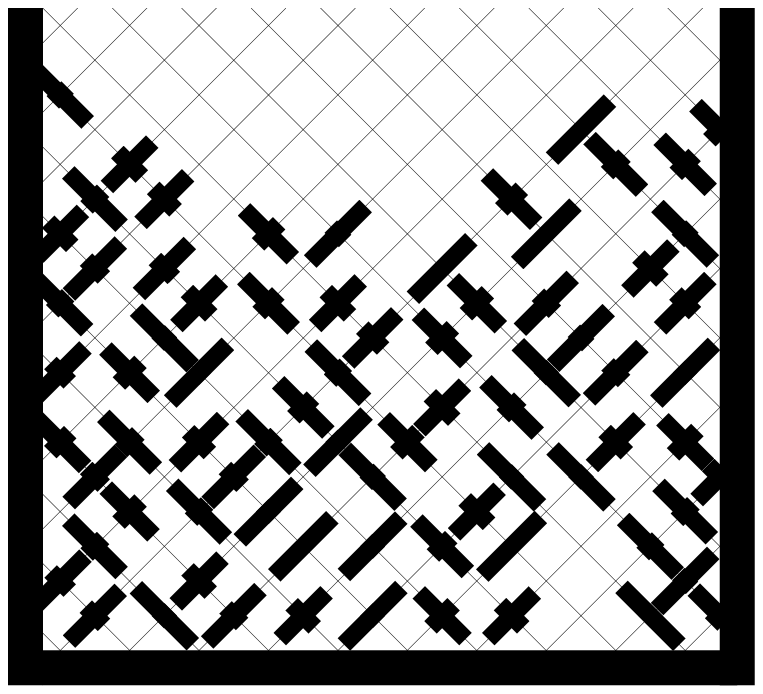,width=6cm,angle=0}
  \hspace{1.0cm}
  \psfig{figure=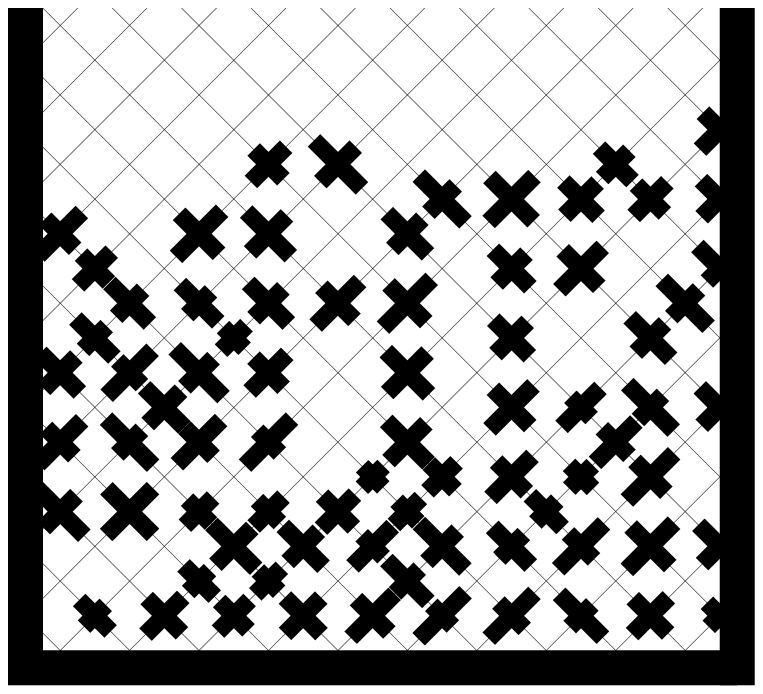,width=6cm,angle=0}
  \vspace{0.5cm}}
        \vspace*{0.1cm}
\caption{Example of a stable packing configuration for a system of 
type $A$ (left) and $B$ (right) (see text). In all our simulations 
we used particles of type $A$.}
\label{conf-bars-sphe}
\end{figure}

In all our simulations we have always used, without
loss of generality, systems with particles of type $A$. 
With the above given rules one can define the
allowed configurations. For instance introducing gravity
one can decide whether a certain configuration (packing)
is mechanically stable or not. For instance 
Fig.~\ref{conf-bars-sphe} shows examples of 
mechanically stable configurations under gravity
for the cases $A$ and $B$. 
In these cases a tilted square lattice was used to implement the existence 
of a preferential direction in the system.

The dynamics will then consist in general in a diffusion
constrained by the particles geometry. In the following sections
we shall be concerned with the case of the vibration-induced 
compaction phenomenon\cite{exp-compaction}.

One of the possible ways to implement a vibration or a shaking
procedure in the RTM model is to consider a Monte-Carlo dynamics
where the particles diffuse on the lattice according to
some rules, as explained below. One can actually choose the 
specific dynamics in several ways.
In the following we discuss some examples.

The system is initialized by filling the container idealized as a lattice
with lateral periodic boundary conditions and a closed boundary
at the bottom. The procedure of
filling consists in inserting the grains at the top of the system,
one at a time, and let them fall down, performing, under the
effect of gravity,
an oriented random walk on the lattice, until they reach
a stable position, say a position in which they cannot fall further.
This filling procedure is realized by the addition of one
particle at a time and stops when no particle can enter the
box from the top anymore.

In the case of shaking the dynamics can be divided in
several alternating steps where the system is perturbed 
by allowing the grains to move in any allowed directions with a 
probability $p_{up}$ to move upwards (with $0 < p_{up} < 1.0$) and a 
probability $p_{down}=1-p_{up}$ to move downwards.
Each step lasts until a fixed number of $N$ moves per particle have 
been attempted with a fixed value of $x=p_{up}/p_{down}$. 
The quantity  $x$ can be related to the adimensional acceleration
$\Gamma$ used in compaction experiments\cite{exp-compaction} through the 
relation $\Gamma \simeq 1 / log(1/\sqrt x)$.

More precisely the single dynamical step consists of the following
operations: 1) extracting  with uniform probability a grain;
2) extracting a possible movement for this grain among
the nearest  neighbours according to the probabilities $p_{up}$ and 
$p_{down}$; 3) move the grain if {\bf all} the possible geometrical 
constraints with the neighbours are satisfied.

One is then free to choose the desired sequence of steps
depending on the quantities one wants to monitor or on the 
experimental procedures one wants to reproduce.

The easiest possibility is the {\em continuous shaking}
realized by letting the system evolve continuously with a fixed 
and constant $x$. We shall discuss this case in Sect. III.
Another possible procedure, explored in \cite{prltetris}, mimics the process
of tapping and it is composed by two alternating steps.
First, in a ``heating'' process (tapping) 
the system is perturbed by allowing the grains to move in any
allowed directions with a probability $p_{up}$ to move upwards 
and a probability $p_{down}=1-p_{up}$ to move downwards.
After each tapping has been completed (i.e. a fixed number of $N$
 moves per particle have been attempted with a fixed value of
$x=p_{up}/p_{down}$) we allow the system to relax setting $p_{up}=0$.
The relaxation process (``cooling'') is supposed to be completed
when no particles can move anymore under the effect of gravity,
i.e. unless $p_{up}$ is switched on.
After this relaxation the system is in a stable static state and one
restarts the cycle.                         


More generally one can define complex cycles where one changes 
the value of $x$ during the evolution of the system according to some 
specific temporal function. This is the most general case that allows 
to impose to the system specific histories. We shall discuss this case 
in Sect. IV.

Before describing in a detailed way the results let us define
the different quantities we shall be monitoring during the
shaking procedure. Denoting with $(i,j)$ the coordinates of a generic site,
where $i$ indicates the horizontal coordinate and $j$ the vertical one
along the direction of gravity,  we can define with $m(i,j)$ the mass 
content of the site $(i,j)$ in such a way that $m(i,j)=1$ if the site 
contains a particle and $m(i,j)=0$ otherwise.
With these definitions in mind we shall monitor the evolution of 
the following quantities:

\begin{itemize}

\item[]{\underline{\em Density Profile}} 
The density profile gives the value of the density (averaged 
over horizontal layers) as a function of the height. It represents 
the simplest, though rough, way to characterize the inhomogeneities 
in the system: since the gravity is acting in the vertical 
direction, we concentrate on the heterogeneities that can occur in this
direction. In formulas we have 
\begin{equation}
p(j,t)= \frac{1}{L} \sum_{i=1}^{L} m_{i,j}(t) \,\,\,\,\,\, \mbox{for} \,\,  j 
\in [0: M]
\label{profile} 
\end{equation}
where $L$ is the width of the system and $M$ its maximal height;

\item[]{\underline{\em Average density}}  
The density of the packing, i.e. the
fraction of sites occupied with respect to the total number of sites,
is measured after each relaxation step and, in correspondence with real
experiments, we plot the behaviour of this density as a function of time.
In order to avoid finite-size effects we considered systems with
a linear size of at least $L=50$ sites and, in order to be sure to observe
bulk effects, we measured the density in the lower $25\%$ or $50\%$ 
of the system;

\item[]{\underline{\em Response function}} 
The response function is defined as the change in the potential energy 
of the system for a small change in the value of $x$. What we do in 
practice is evolve the system for a certain time $t_w$. At $t_w$ 
we define a replica of the system whose mass content will be described 
by $m_{i,j}^{r}$ and we let the original system evolve with the same 
$x$ and the replica with $x^{\prime} = x + \delta x$ where $\delta x$ 
is small enough to keep us in the linear response approximation. 
At times larger than $t_w$ we monitor the evolution of the potential 
energy for the two replicas defined as 
$P(t+t_w) = \sum_{i,j} m_{i,j}(t+t_w) \cdot (j+1)$
and $P_r(t+t_w) = \sum_{i,j} m_{i,j}^{r}(t+t_w) \cdot (j+1)$.
The response of the system is defined as 

\begin{equation} 
R(t+t_w,t_w) =\frac{P_r(t+t_w)-P(t+t_w)}{N_{part}}
\label{response}
\end{equation}
where $N_{part}$ is the total number of particles in the system. We also
monitor the density profile of the replica, $p^r(j,t+t_w)$,
and the difference $\Delta p(j,t+t_w)=p^r(j,t+t_w)-p(j,t+t_w)$ between
the profiles.
Note that $R(t+t_w,t_w)$ can also be written as
$R((t+t_w,t_w)=\sum_j \Delta p(j,t+t_w) \cdot (j+1) \cdot L/N_{part}$.

\item[]{\underline{\em Correlation functions   }} We have 
considered two kind of measures of the temporal correlations. 
The two times mass-mass correlation function, defined as:
\begin{equation}
C(t+t_w,t_w)=
\frac{1}{N_{part}} \overline{\sum_{i,j} (m_{i,j}(t+t_w) \cdot m_{i,j} (t_w))}
\label{eq:defcorr}
\end{equation}
and the mean-square distance between the potential energies
at times $t_w$ and $t+t_w$, defined as:
\begin{equation}
B(t+t_w,t_w)= \overline{\left(\frac{\sum_{i,j} (j+1) 
\cdot m_{i,j}(t+t_w)}{N_{part}} - 
\frac{\sum_{i,j} (j+1) m_{i,j} (t_w)}{N_{part}}\right)^2}=
\overline{\left(\overline{h}(t+t_w)-\overline{h}(t_w)\right)^2}
\end{equation}
where the overbar indicates the average over  different realizations and
$\overline{h}(t)$ indicates the height of the center of mass at time $t$.

\end{itemize}

It is worth noticing that we decided to measure the two-times 
mass-mass correlation function via eq.~\ref{eq:defcorr} instead of the
the two-times correlation function for the global density 
\cite{nicodemi_coniglio} because, as it will be clear in the following,
the definition of the global density is somehow arbitrary in these 
systems with a preferential direction imposed by gravity.

We have used system sizes $L \times M$
of $60 \times 60$, $120 \times 60$ and
$60 \times 120$ to ensure that finite-size effects were irrelevant.
We let the system evolve for $10^6$ Monte-Carlo steps per particle
after a preparation (waiting at constant $x$ or other more complicated
histories) during up to $10^6$ Monte-Carlo steps per particle. Averaging 
was performed on up to $1000$ samples.

\section{Response properties of a system subject to continuous shaking}

In this section we present the results of the simulations 
for a system subject to continuous 
shaking at constant $x$. 

\begin{figure}[h]
\centerline{
       \psfig{figure=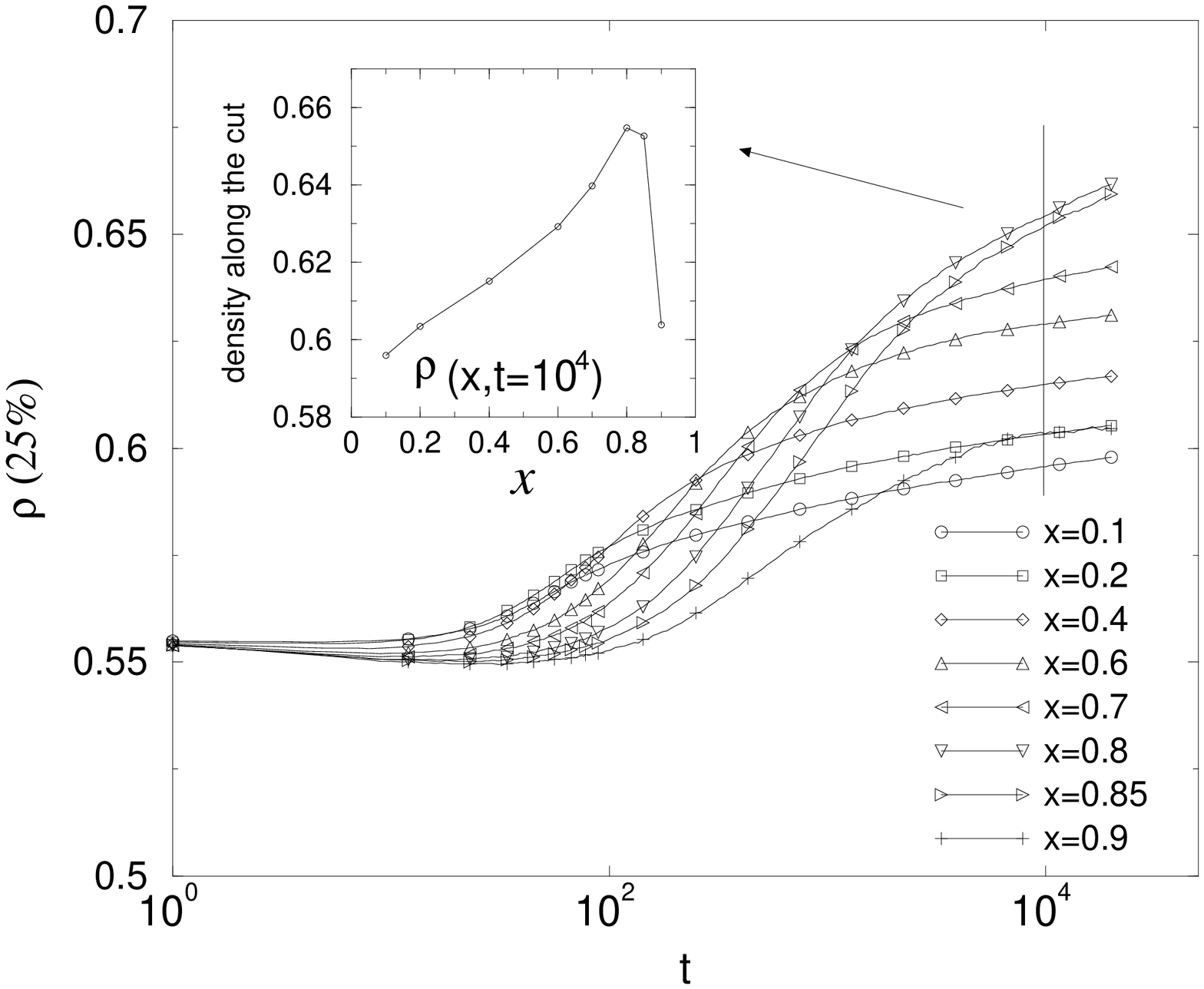,width=6cm,angle=-90}
       \hspace{1.0cm}
       \psfig{figure=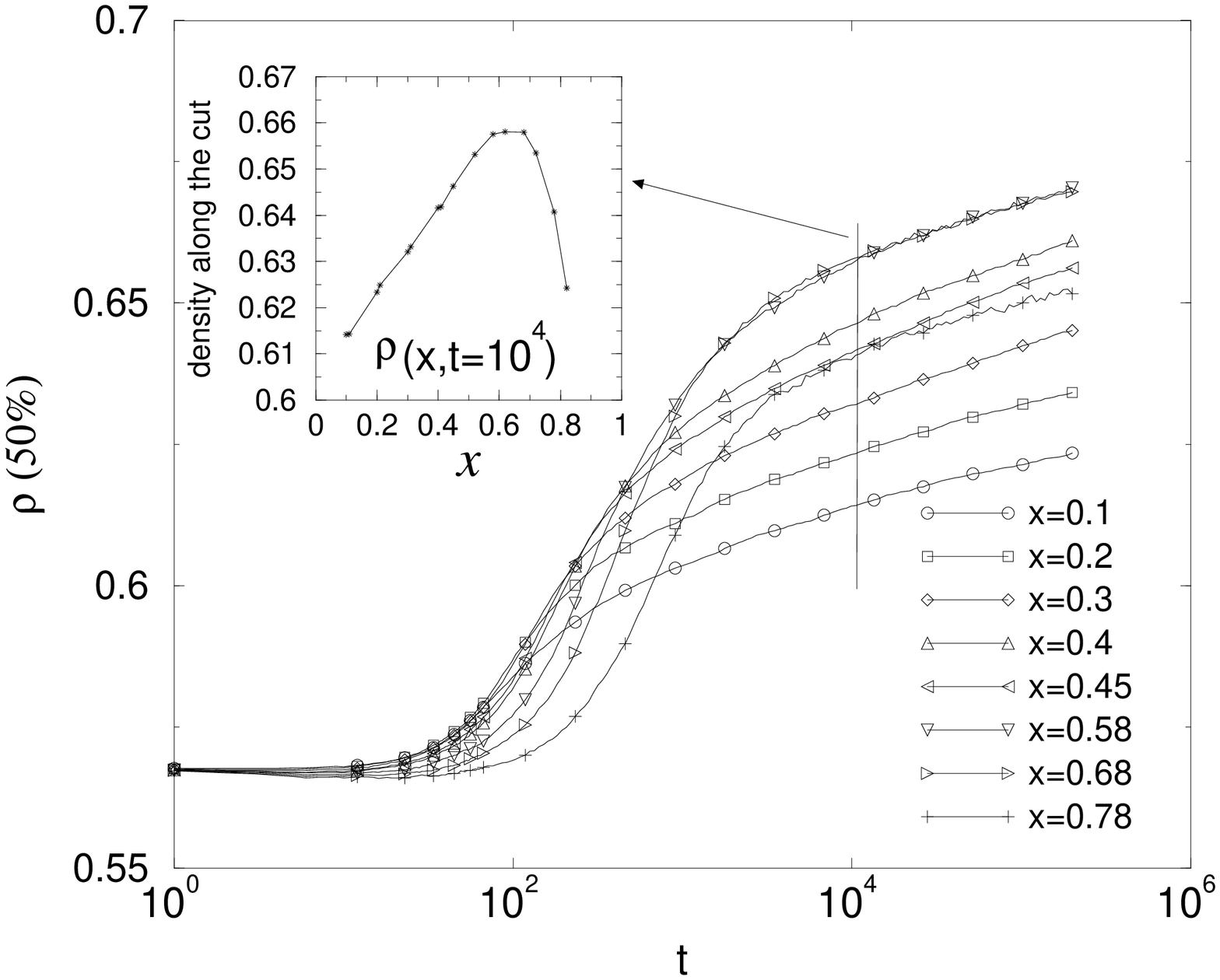,width=6cm,angle=-90}
  \vspace{0.5cm}}
\caption{Time-evolution of the bulk density (lower $25\%$ (left) and $50\%$
(right) of the system) for several shaking amplitudes.
The insets show the density at time $t=10^4$ as a function of $x$. 
Note in both cases the existence of an optimal value of $x$, that
depends on the fraction of the system where one measures the average 
density.}
\label{fig:rho}
\end{figure}

The mean density, calculated in the lower $25\%$ and $50\%$ of the box,
grows slowly, from an initial value $\rho_0$.
This value is not universal but depends on the choice of the particles
and on the fraction of the system where one performs the average.
Fig.s~\ref{fig:rho} show the evolution of the density measured in the 
lower $25\%$ and $50\%$ of the system for different values of $x$.
It is interesting to notice that there exists an optimal value of $x$
that allows to obtain the maximal density in the fraction of the 
system considered for the averaging procedure. At small times 
this optimal value depends on time due to the crossing of the different 
curves. This fact already suggests that, in order to compactify the system, 
the optimal strategy will not be to keep $x$ constant, but rather to vary it. 
This type of behaviour was also recently observed in the parking lot model
\cite{viot}. However, in our case the mechanism responsible for these
phenomena can be understood looking at the density profiles (see below),
a concept which is absent in the parking model.
Moreover if we compare the data for the bulk density 
computed as $25\%$ or $50\%$ of the bulk, we see that 
the curves differ and that the optimal value of $x$ is
around $x\simeq 0.6$ in the first case, and around $x\simeq 0.8$ in the
other. This feature can be explained by noticing the combined effect of two 
factors: on the one hand the bulk compactifies better for larger $x$: 
for example, in the data of \cite{exp-compaction}, 
taken at the bottom of the sample, the density is an increasing function 
of the shaking amplitude; on the other hand, increasing $x$, the interface 
becomes broader and can affect the global bulk density effectively 
reducing it. As a result there will be an optimal $x$ which for instance 
is smaller considering the measure of the density in the lower $50\%$
of the system with respect to the measure in the lower $25\%$ because 
one needs a larger $x$ in order to extend the interface deeper and 
deeper.
This phenomenology represents an indication that heterogeneities 
are quite relevant and that particular attention needs to be taken 
in defining the observed quantities.
Our finding of an optimal shaking amplitude is in fact 
in agreement with previous experimental results~\cite{exp-compaction}.
We recall for this that $x$ and the adimensional acceleration $\Gamma$ 
can be linked by the relation $\Gamma \simeq 1 / log(1/\sqrt x)$.
In compaction experiments ~\cite{exp-compaction} it was observed that 
the density, measured on the lowest part of the system,
was an increasing function of the shaking amplitude.
However, the authors were expecting this density to ``decrease upon
further acceleration increase'', since their experiments probed only
the regime of relatively low shaking intensity. Moreover, such a
decrease is expected to be more apparent in two-dimensional systems
\cite{exp2d}, which is the case of our model.

\begin{figure}[h]
\centerline{
       \psfig{figure=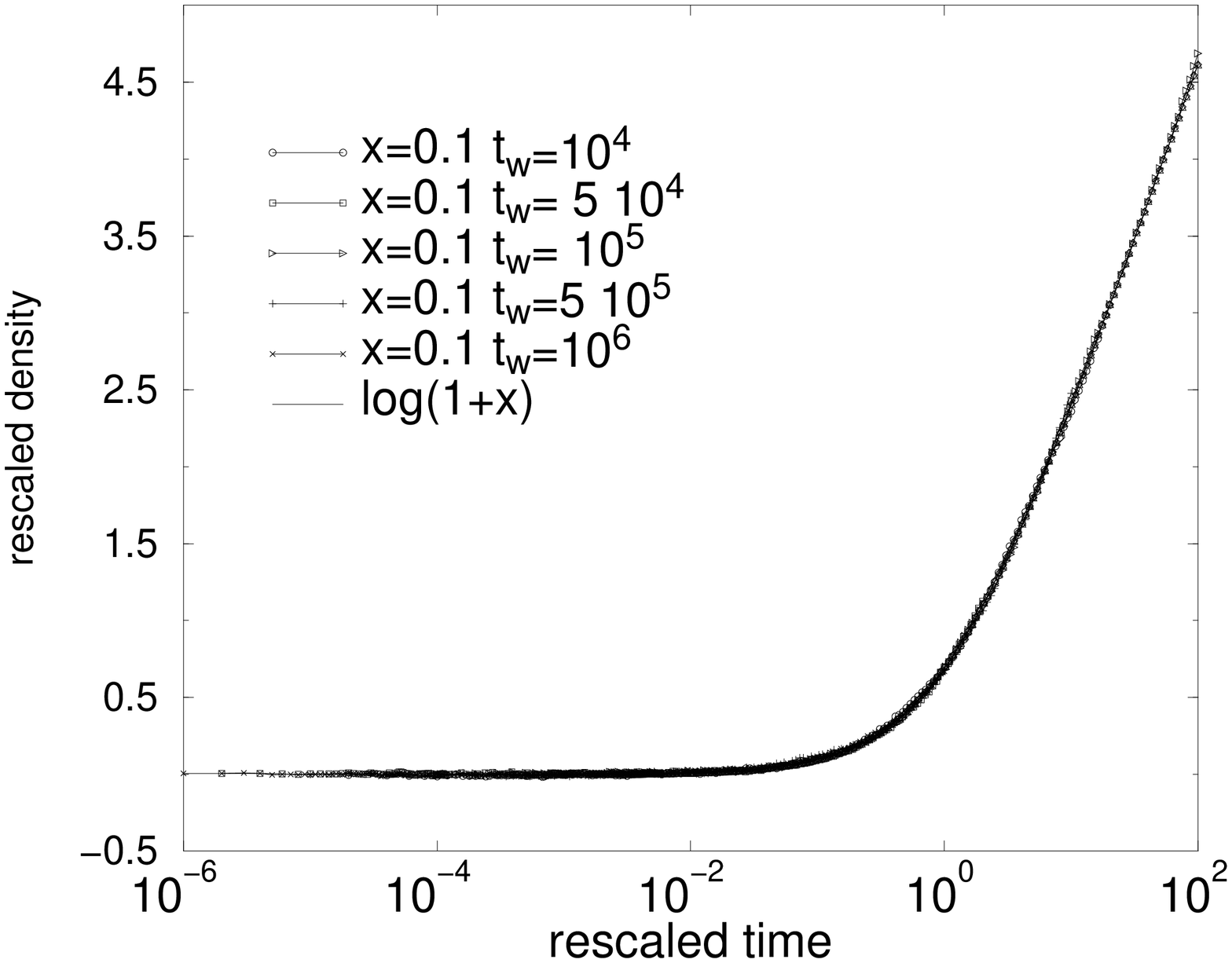,width=5cm,angle=-90}
       \psfig{figure=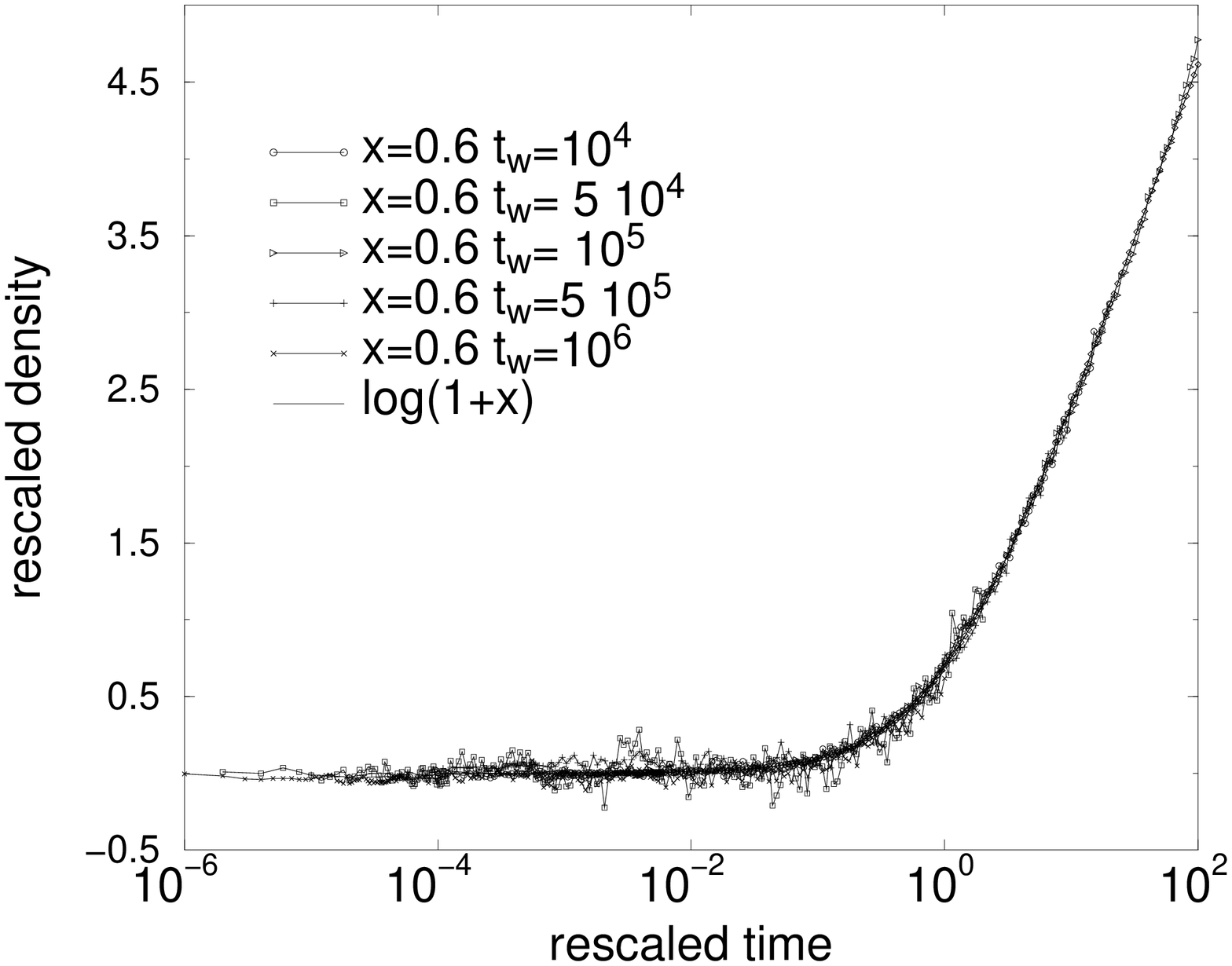,width=5cm,angle=-90}
       \psfig{figure=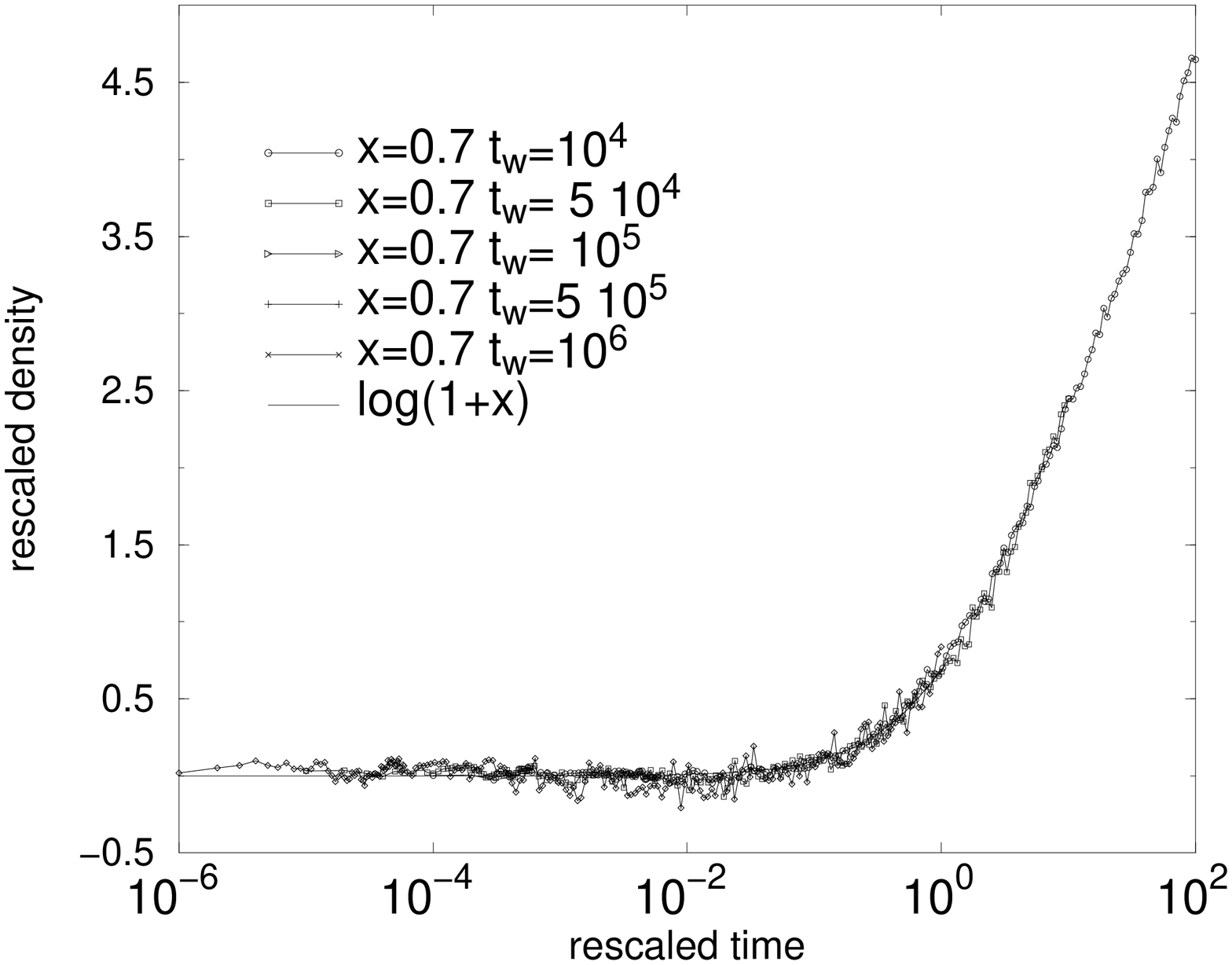,width=5cm,angle=-90}
         \vspace{0.5cm}}
\caption{Collapses of the rescaled density 
$(\rho (t) - \rho_0)/(B(\rho_\infty - \rho (t)))$ versus the rescaled
time $\tilde{t}=t/\tau$, where $B$, $\rho_\infty$ and $\tau$ are fit
parameters, for $x=0.1$ (left), $x=0.6$ (center) and $x=0.7$ (right),
and for densities measured after various $t_w$.}
\label{fig:rho1}
\end{figure}

The growth law for the density has been shown by various authors to be,
experimentally\cite{exp-compaction} and for many models
\cite{prltetris,mod-compaction}, well described by the functional form:
\begin{equation}
\rho (t) = \rho_\infty - \frac{\rho_\infty - \rho_0}{1+B \ln 
(1+t/\tau)}
\label{eq:rho}
\end{equation}
where $\rho_\infty$, the asymptotic density, $B$ and $\tau$ 
are fit parameters depending on the shaking amplitude.
Also in our case eq.~\ref{eq:rho} is satisfied and one can 
actually collapse all the curves in a unique logarithmic function. 
Fig.~\ref{fig:rho1} shows the collapse of the density curves
measured in the lower $50\%$ of the system (one gets similar results 
using the curves of the density measured in the lower $25\%$ of the system) 
for three different values of $x$ ($x=0.1,\ 0.6,\ 0.7$) and for different 
values of $t_w$ (i.e. the measure of the density begins at $t_w$ 
instead of $0$, which of course changes the parameters in eq. (\ref{eq:rho})): 
the functional form is tested by plotting the rescaled
density $(\rho(t) - \rho_0)/(B(\rho_\infty - \rho (t)))$ versus
$\tilde{t}=t/\tau$, and comparing it to $\ln(1+\tilde{t})$.

In all cases the functional form (\ref{eq:rho}) is quite 
well satisfied with $\tau$ simply proportional to $t_w$, 
$\rho_\infty$ being a bell-shaped function of $x$ \cite{nota2} and $B$ 
exhibiting a complex dependence on $t_w$ and $x$. 
Several remarks are in order. The fitting parameters depend on the fraction
of the system where one measures the density. On the other hand the
logarithmic fit starts to be valid after a transient time of the order of 
$50-500$ iterations depending on $x$ and on the fraction of the system 
used for the measurements. This behaviour is quite understandable.
Before compaction can start in the bulk a weak decompaction is in 
order which is stronger for higher $x$ and which
extends at longer times if one is measuring the density in a deeper
region. One can actually notice the decompaction process in 
Fig.~\ref{fig:rho} especially for large values of $x$. 
Only after the decompaction process ends one can hope 
eq.~\ref{eq:rho} be valid. This is the reason why we have 
considered the collapse of the density curves obtained after a suitable 
waiting time $t_w$.

\begin{figure}[h]
\centerline{
       \psfig{figure=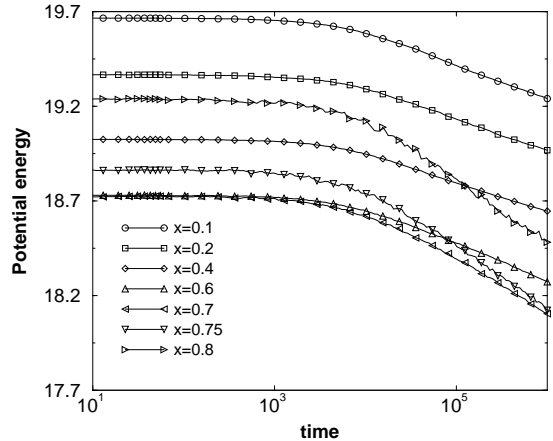,width=6cm,angle=-90}
  \vspace{0.5cm}}
\caption{Potential energy (or mean height) of the system,
for various $x$ and measured after $t_w=10^4$.}
\label{fig:hh}
\end{figure}

A different way of measuring the compaction is to look
at the mean height, or potential energy $P$ defined in section II.
In this case, the whole system is considered. We see in 
Fig.~\ref{fig:hh} that, also in this case, there exists an optimal value
of $x$ for which one has the lowest position of the center of mass.
Also in this case we find an agreement with the experimental findings of 
Ref.~\cite{exp-compaction} where the density measured from the height
of the system was an increasing function of the shaking amplitude, while
the authors expected the presence of a peak ``upon
further acceleration increase''.

It is however already clear how the previous measurements, which
are averages over the whole (or over an extensive part) of the system,
cannot represent a comprehensive information about the system. 
For example, in the measure of the mean height the interface gives a large
contribution while it is practically negligible if one is interested
in bulk properties. 
Therefore before turning our attention to a more detailed description 
of the system let us now study the two-times
correlation function $C(t+t_w,t_w)$ which,
as usual in ageing phenomena\cite{review_aging}, 
gives a determination of the age of
the system, since it depends on $t$ and on $t_w$.

\begin{figure}[h]
\centerline{
       \psfig{figure=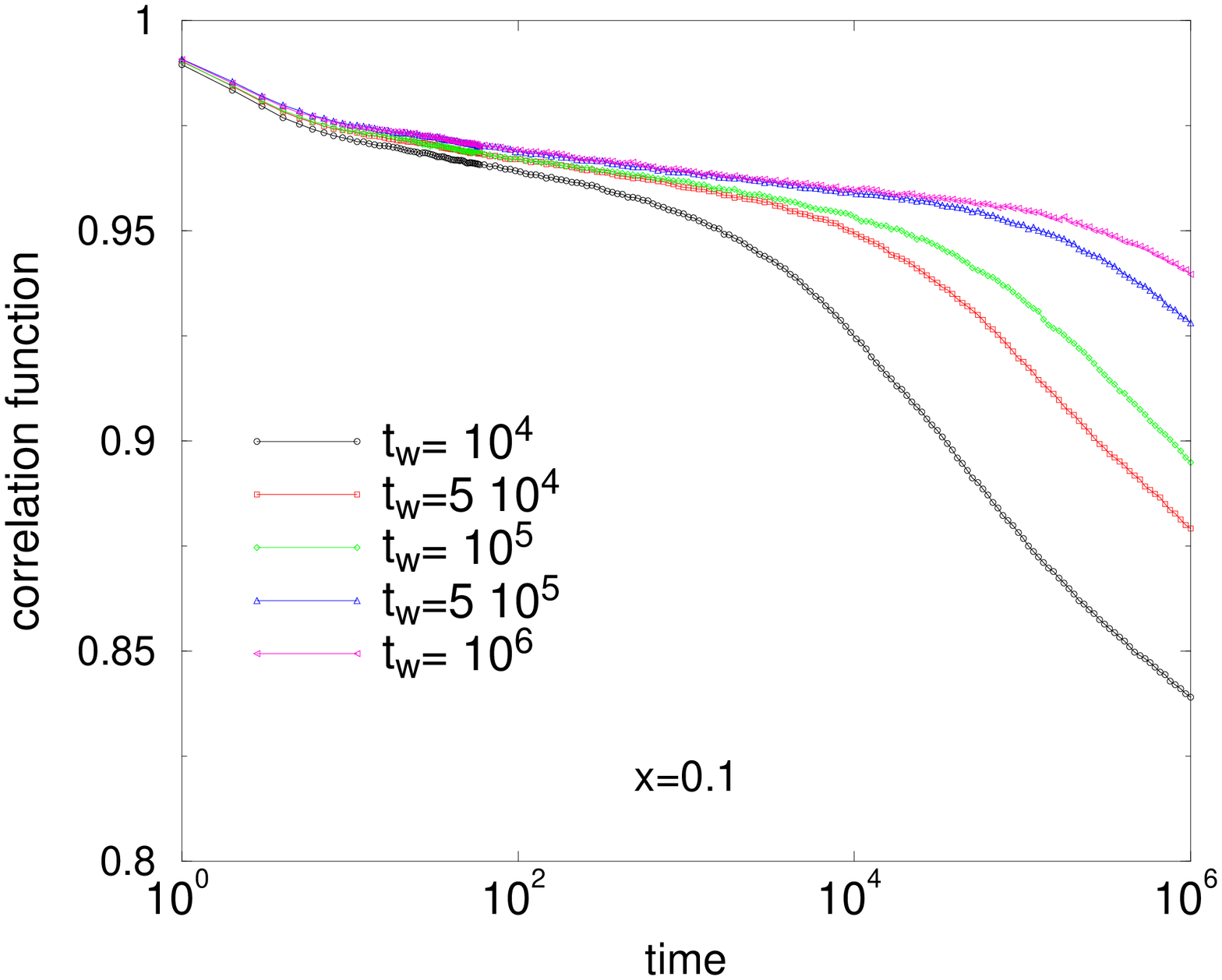,width=5cm,angle=-90}
       \psfig{figure=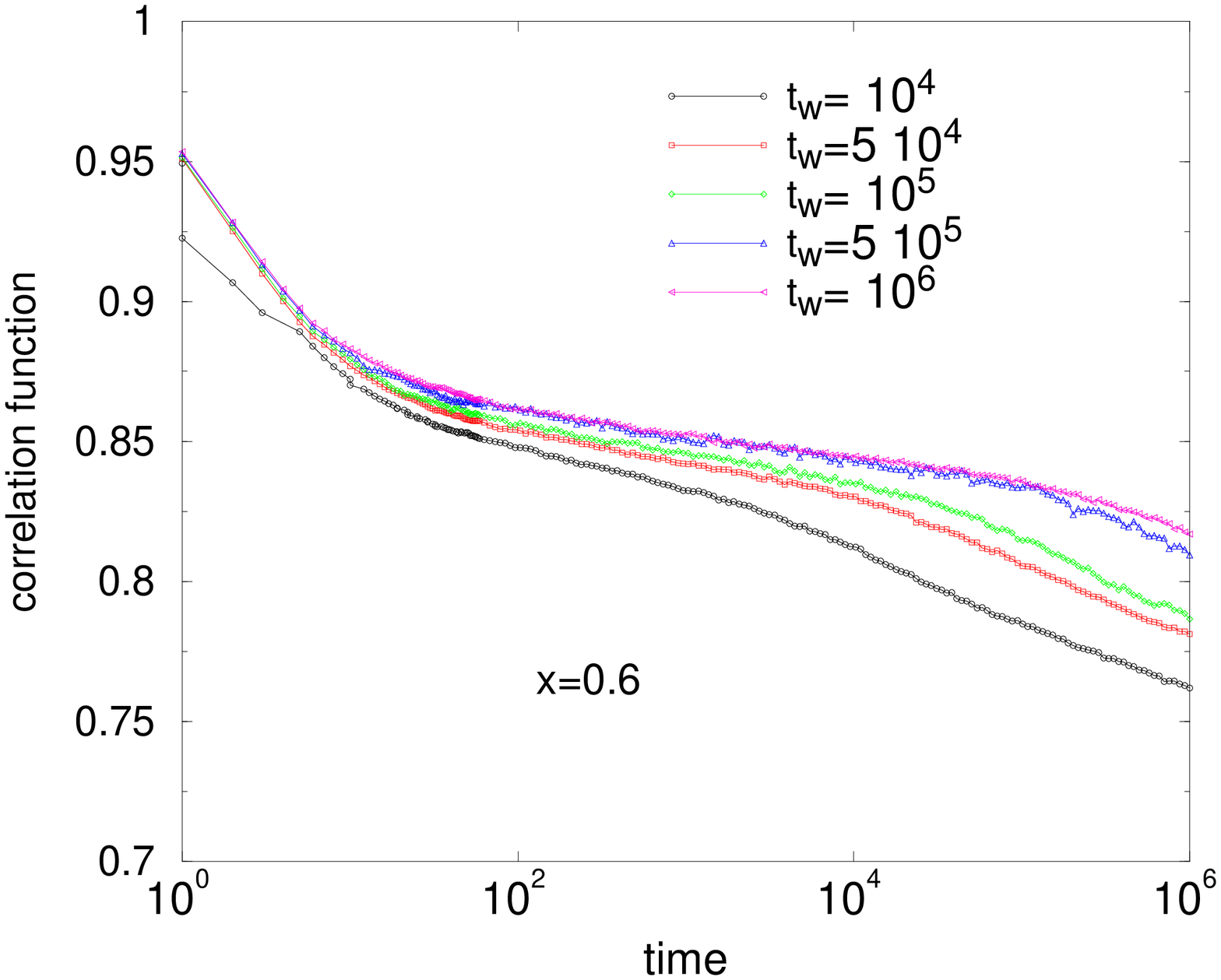,width=5cm,angle=-90}
       \psfig{figure=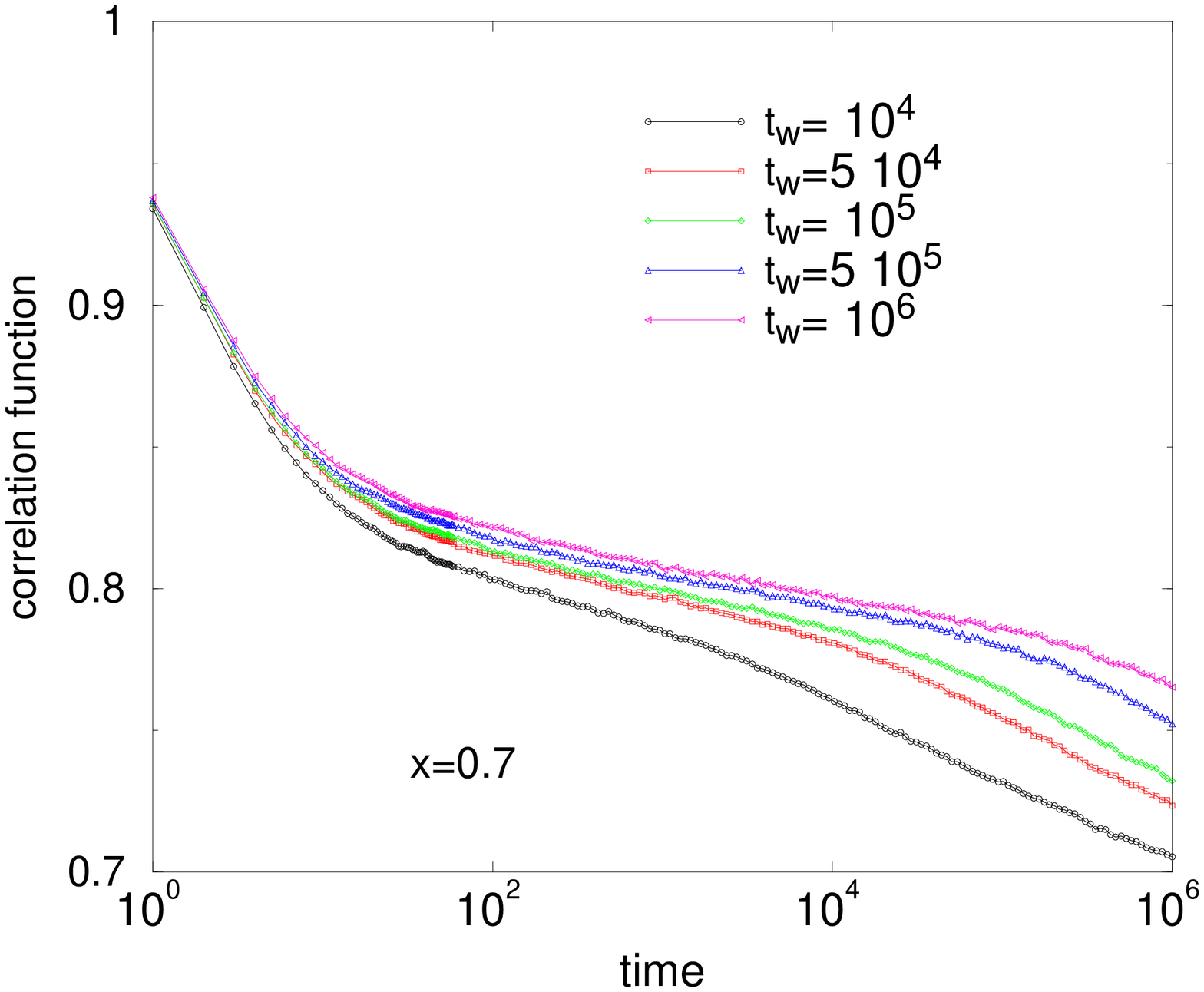,width=5cm,angle=-90}
        \vspace{0.5cm}}
\caption{Correlation function $C(t+t_w,t_w)$ as a function of
$t$, for $x=0.1$ (left), $x=0.6$ (center) and $x=0.7$ (right), 
and various values of $t_w$. The ageing behaviour with
a two-steps relaxation is evident.}
\label{fig:corr}
\end{figure}

\begin{figure}[h]
\centerline{
       \psfig{figure=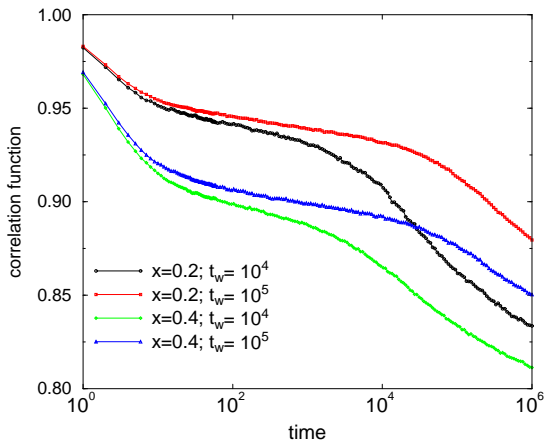,width=6cm,angle=-90}
         \vspace{0.5cm}}
\caption{Correlation function $C(t+t_w,t_w)$ as a function of
$t$, for $x=0.2$ and $x=0.4$, $t_w=10^4$ and $t_w=10^5$: the correlation
decays faster for higher $x$.}
\label{fig:corr2}
\end{figure}

In Fig.s~\ref{fig:corr} and \ref{fig:corr2} we show the behaviour of 
$C(t+t_w,t_w)$ for several values of $x$ and $t_w$. 
We observe for this model the typical ageing behaviour,
with a first part for $t \ll t_w$
approaching a quasi-equilibrium curve where time-translation
invariance is respected (i.e., for $t \ll t_w$,
$C(t+t_w,t_w)$ approaches a curve depending only on $t$)
a plateau, and, at $t \gg t_w$, a second decay, dependent on
$t_w$, corresponding to ageing.
We note that this behaviour, very common in ageing phenomena 
\cite{review_aging}, is more realistic than the one of the simplest
Tetris model\cite{prltetris}, where the plateau is at $C=1$, 
so that only the second decay is observed \cite{nicodemi_coniglio}. 
As $t_w$ grows, the plateau is better defined and the second decay 
appears at longer times.

The global properties of the correlation function depend smoothly
on the shaking amplitude: as $x$ grows, the correlation decays
faster, as shown in Fig.~\ref{fig:corr2}.
However, the curves do not differ qualitatively and have similar
shapes.

In \cite{nicodemi_coniglio} the two-times correlation function 
for the global density was measured in the simplest version of 
the Tetris model. In this case it was proposed that the relaxation 
was of the form $\ln(t)/\ln(t_w)$ (the exact proposed form was 
$(1-c_\infty) \ln((t_w+t_s)/\tau)/\ln((t+t_w+t_s)/\tau) + c_\infty$, 
where $c_\infty$, $t_s$ and $\tau$ were three fit parameters),
for very small values of $x$ (in the range $[10^{-4};10^{-1}]$). This
form could fit all the curves since, as already mentioned, the first
decay was inexistent.

\begin{figure}[h]
\centerline{
       \psfig{figure=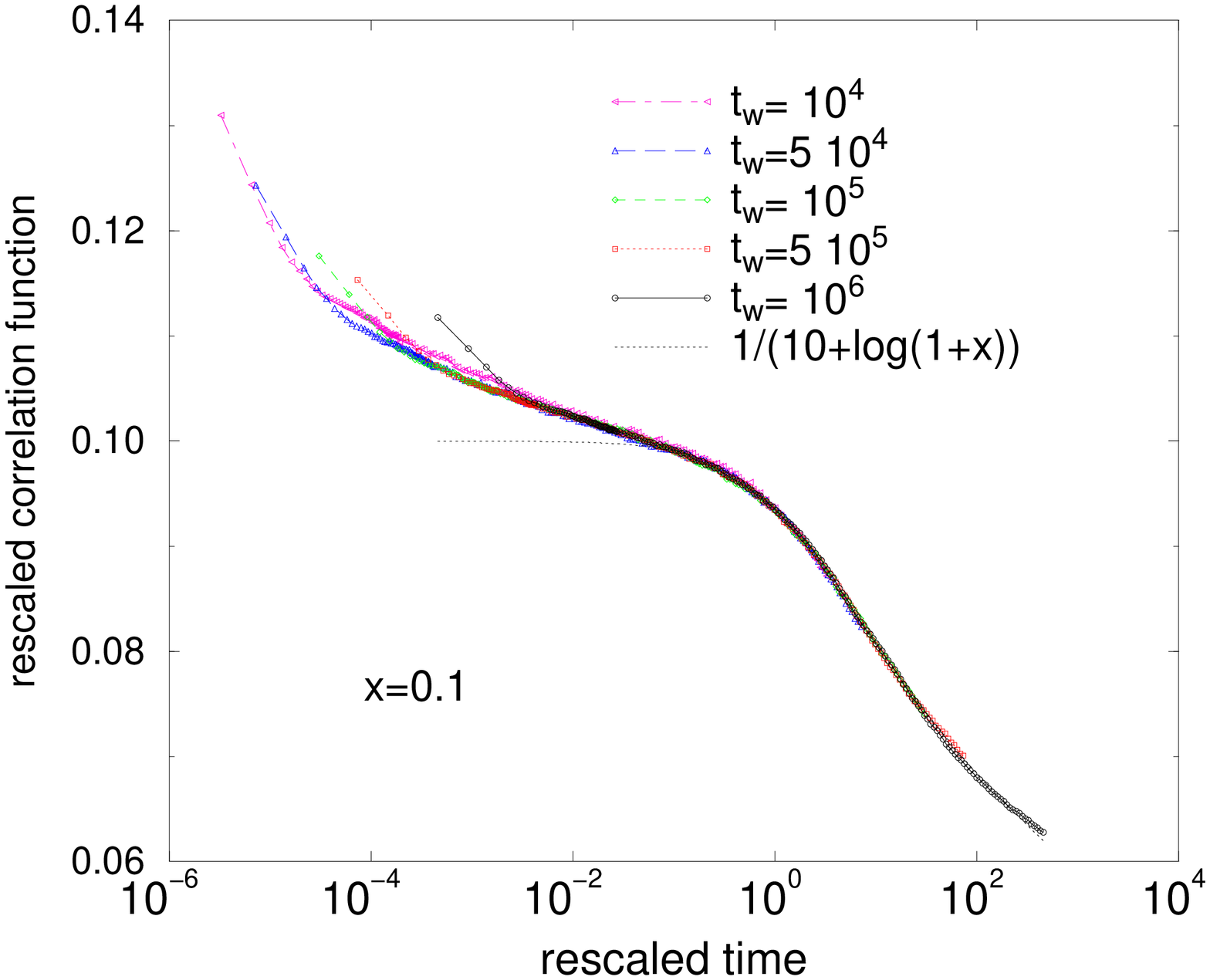,width=5cm,angle=-90}
       \psfig{figure=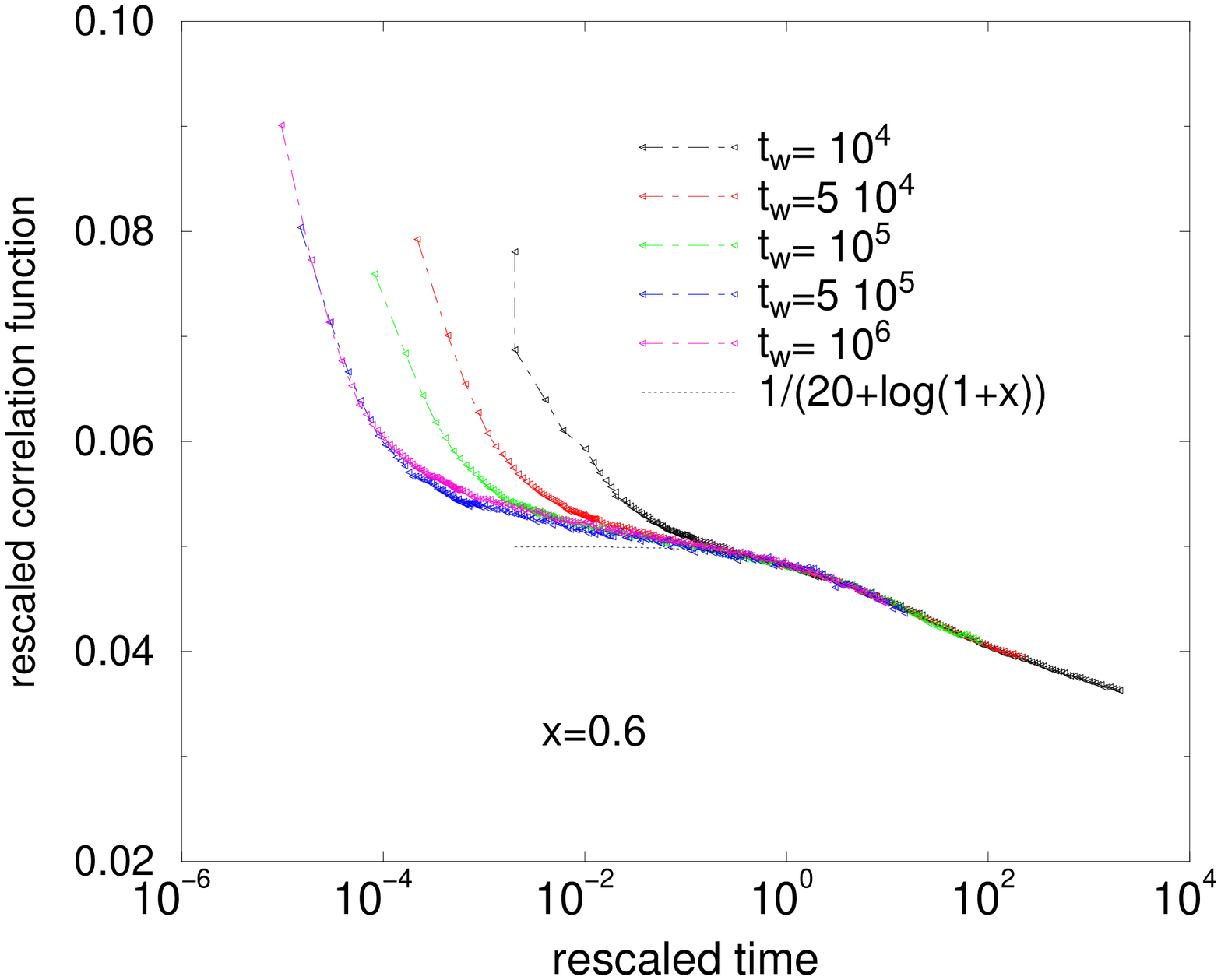,width=5cm,angle=-90}
       \psfig{figure=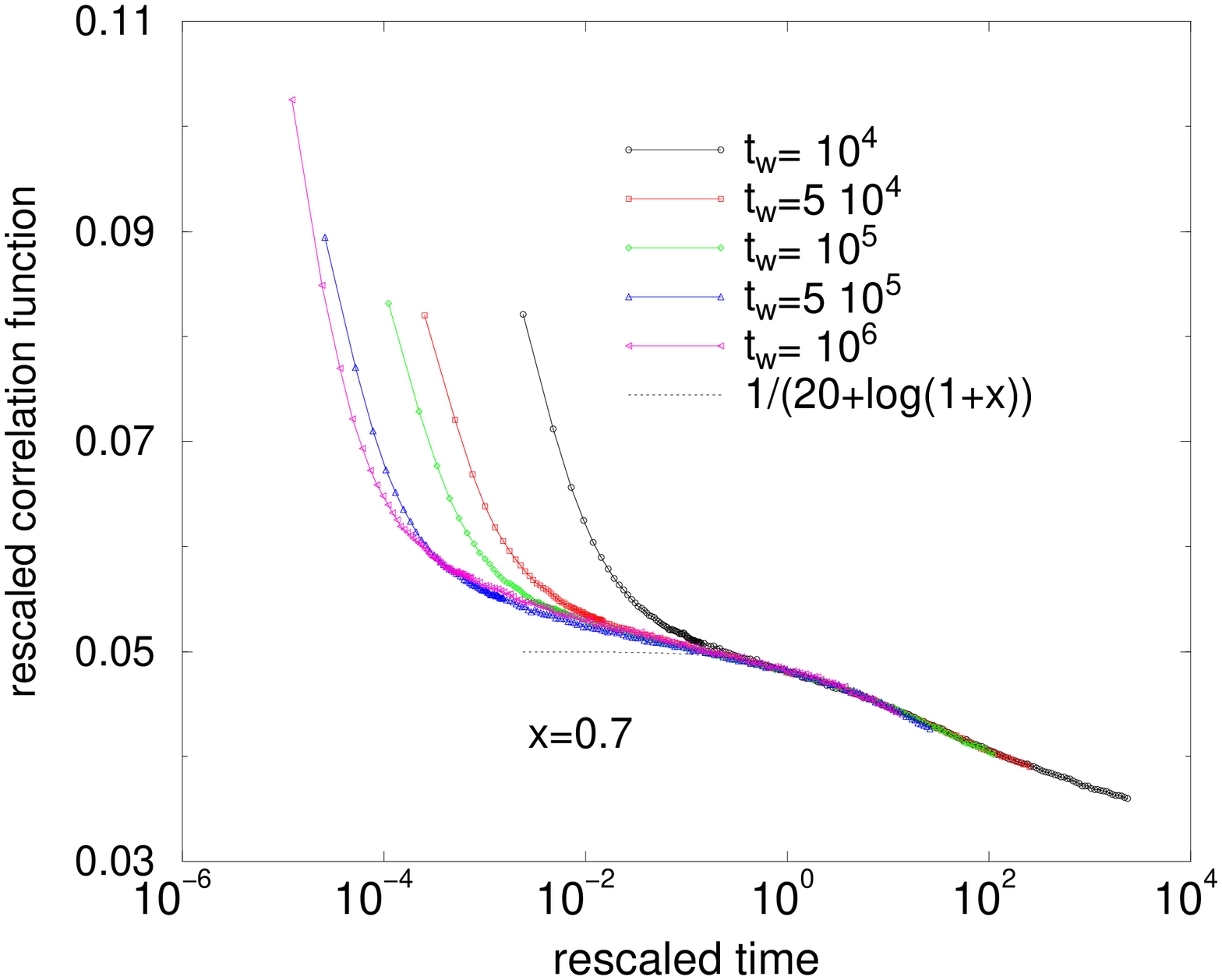,width=5cm,angle=-90}
  \vspace{0.5cm}}
\caption{Collapse of the ageing part of the curves for the same values of
$x$ and $t_w$ ($5$ values for each $x$) as in Fig.~\ref{fig:corr}; 
the rescaled correlation (see text) is plotted versus the
rescaled time $t/\tau$, and compared to the form
$ 1/(A + \ln (1+\tilde{t}))$; the
first part of the curves, not collapsed, corresponds to the first
relaxation (approach to a quasi-equilibrium behaviour).}
\label{fig:corr_coll}
\end{figure}

\begin{figure}[h]
\centerline{
       \psfig{figure=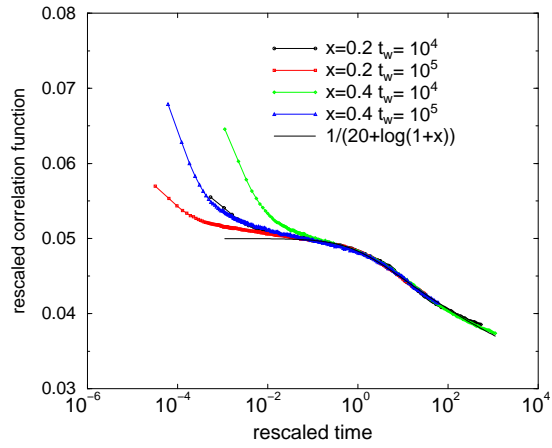,width=6cm,angle=-90}
         \vspace{0.5cm}}
\caption{As in Fig.~\ref{fig:corr_coll} for $x=0.2$ and $x=0.4$.}
\label{fig:corr_coll2}
\end{figure}

Since, as it has been pointed out in the above discussion, 
the definition of the system density is somehow arbitrary, we decided 
to measure the two-times mass-mass correlation function 
via eq.~\ref{eq:defcorr}.
In our case, we only attempt to fit the second decay, i.e. the ageing part;
since we expect weak ergodicity breaking \cite{bouchaud}, we propose
a form going to zero at long times:
\begin{equation}
C(t+t_w,t_w)= \frac{a}{1 + b \ln (1+t/\tau)} \ \mbox{for}\ t \gg t_w
\label{eq:corr}
\end{equation}
with $a$, $b$, $\tau$ fit parameters.
We show in Fig.s~\ref{fig:corr_coll} and ~\ref{fig:corr_coll2}
that we can collapse the curves using eq.~\ref{eq:corr} and
plotting the rescaled correlation function $\tilde{C}$ versus 
the rescaled time $\tilde{t}$ according to
\begin{equation}
\left\{
\begin{array}[c]{ccc}
\tilde{C} &=& \frac{b C}{a + (Ab -1)C} \\
\tilde{t} &=& t/\tau,\\
\end{array}
\right.
\end{equation}
and obtaining $\tilde{C} = 1/(A + \ln (1+\tilde{t}))$.
$A$ is an irrelevant parameter introduced to avoid the divergence
of $\tilde{C}$ at $\tilde{t} \to 0$, and whose value can be read in 
the figures.
The behaviour of the fit parameters is the following: $a$ is roughly
constant, while $\tau$ evolves proportionally to $t_w$, and
$b$ as the inverse of the logarithm of $t_w$. This leads to
the conclusion that the overall behaviour of 
$C(t+t_w,t_w)$ is of the form $log(t_w+t)/log(t_w)$. 
We note that this is in agreement with the findings of 
\cite{nicodemi_coniglio}, on a much wider range for $x$, 
but in contrast with the parking-lot model, for which a $t/t_w$ behaviour
has been observed \cite{viot_private}, where, as in 
\cite{nicodemi_coniglio}, a different definition of
the correlation function was used: the correlation
function of the densities. An experimental measure of the correlation
functions would be welcome to discriminate between these predictions.

\begin{figure}[h]
\centerline{
       \epsfig{figure=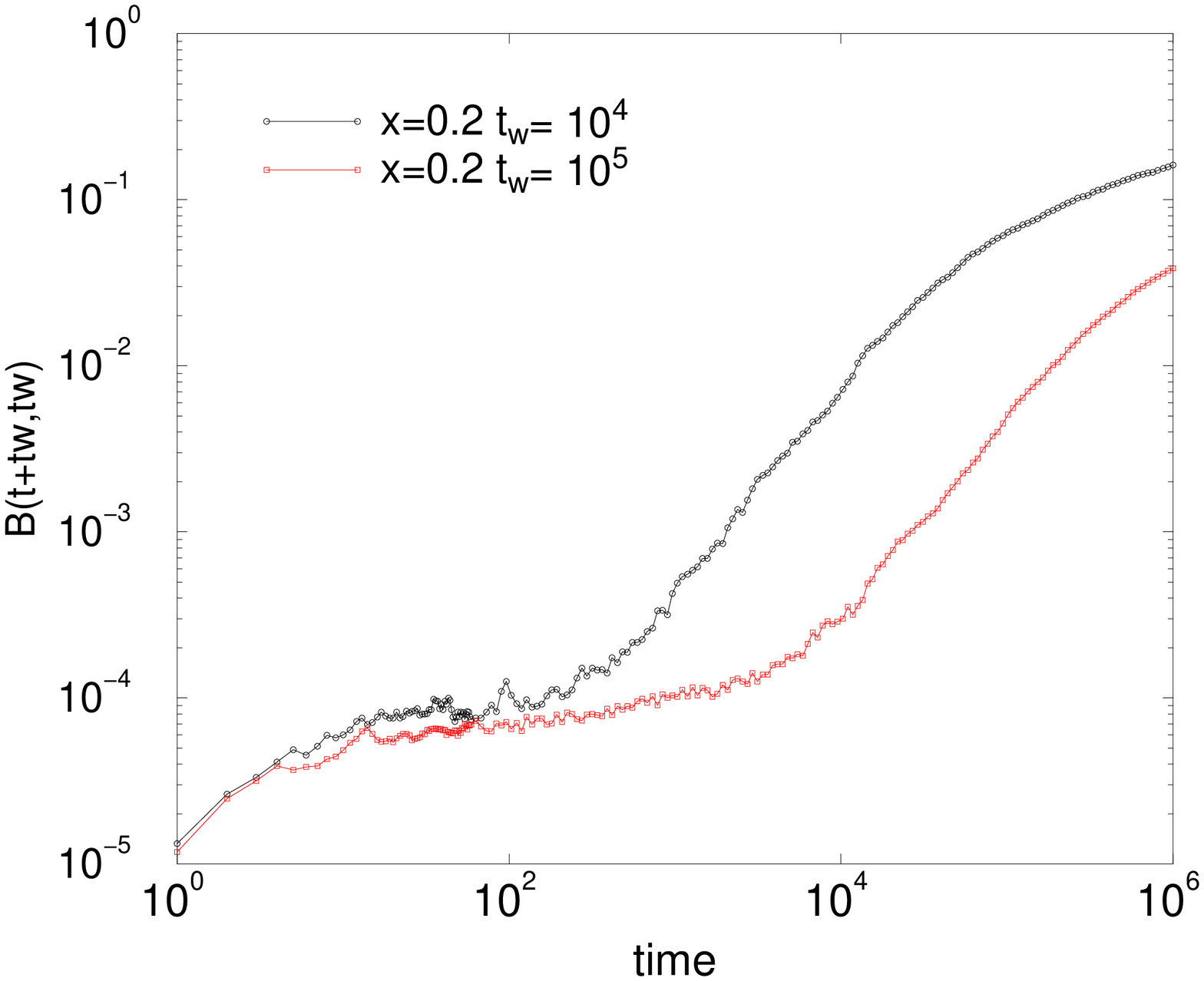,width=5cm,angle=-90}
       \epsfig{figure=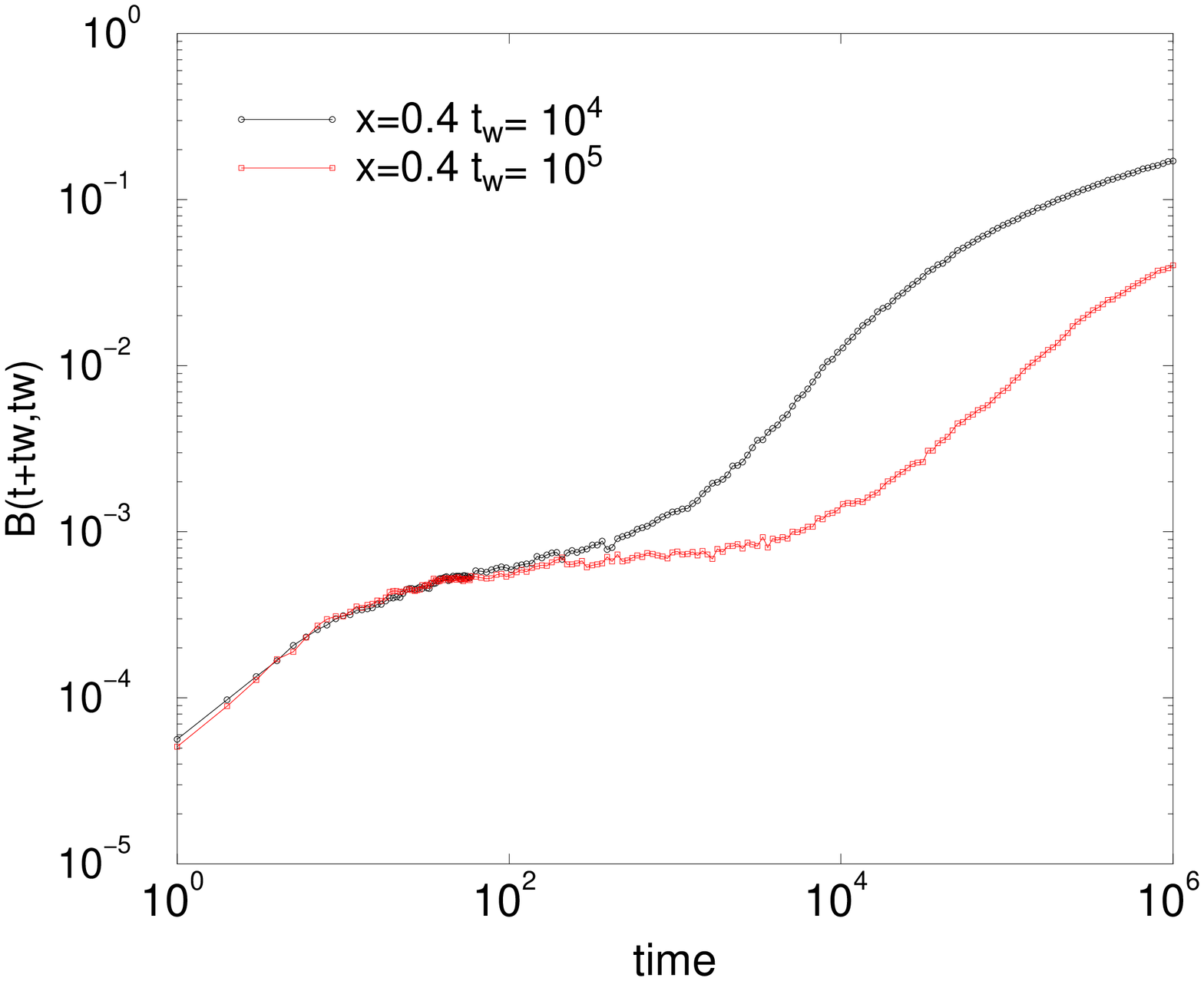,width=5cm,angle=-90}
       \epsfig{figure=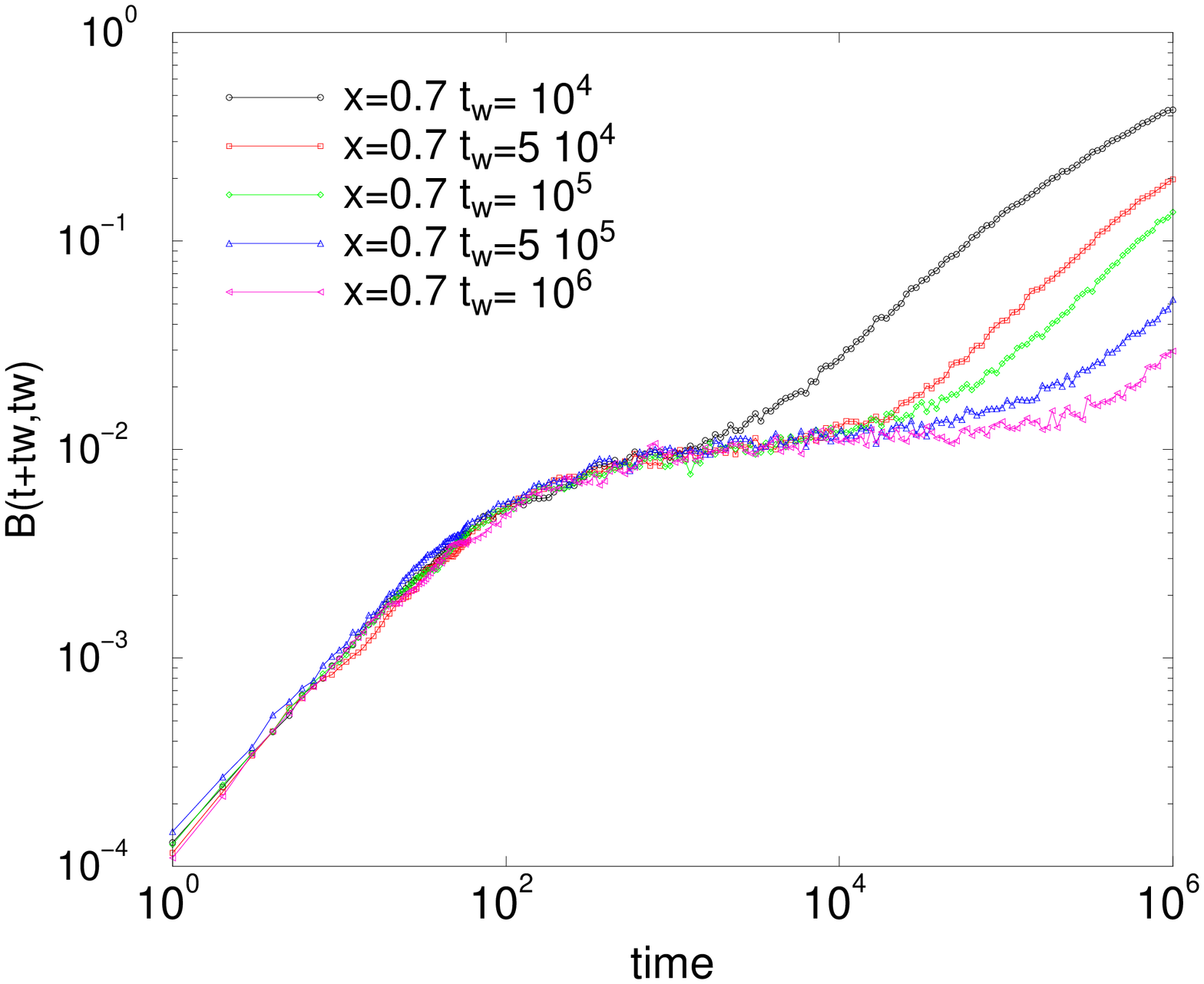,width=5cm,angle=-90} 
  \vspace{0.5cm}}
\caption{Evolution of the mean-square distance between the potential 
energies, $B(t+t_w,t_w)$, with $t$, for 
$x=0.2$ (left), $x=0.4$ (center) and $x=0.7$ (right), and various $t_w$.
         }
\label{fig:B}
\end{figure}

For completeness, we show in Fig.~\ref{fig:B} the behaviour of the
mean-square distance between the potential energies, $B(t+t_w,t_w)$
defined in the introduction though eq.~\ref{eq:corr}. 
$B(t+t_w,t_w)$ is an increasing function of $t$, 
displaying ageing behaviour with two steps separated by a plateau,
as for the correlation function.

Before turning our attention to the response function
defined in eq.~\ref{response} let us add one important information
concerning the density profiles. It turns out in fact that it is essential
to look at the inhomogeneities in the system in order to correctly interpret
the response results. We shall do this looking at the density profiles
and at the differences between the profiles of the system and its
perturbed replica, which will give information on the spatial
structures, at least along the vertical direction.

\begin{figure}[h]
\centerline{
       \psfig{figure=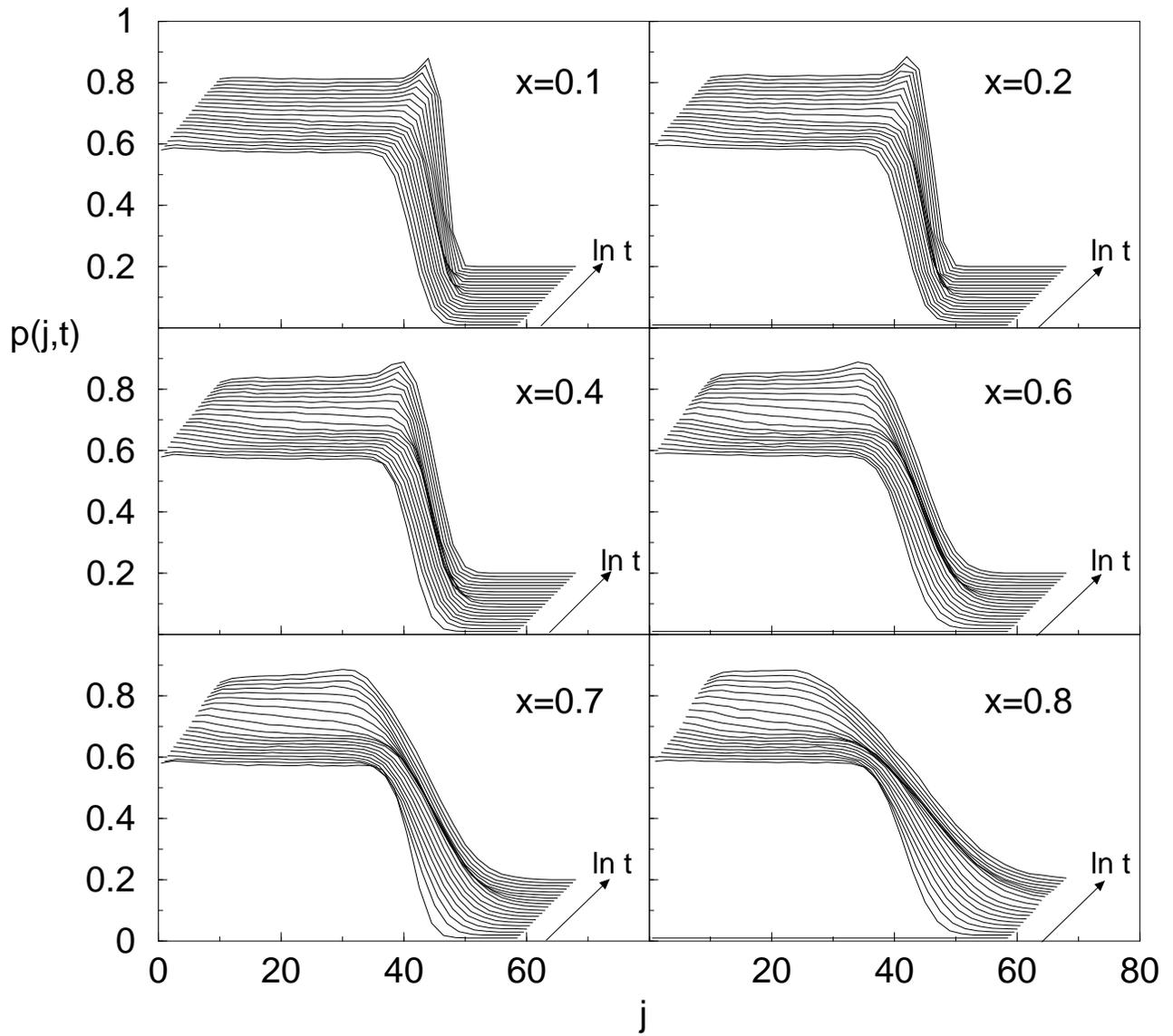,width=15cm,angle=-90} 
  \vspace{0.5cm}}
\caption{Density profiles for various values of $x$, evolving
in time ($t=0$ to $t=10^4$),
with no waiting time: the evolution
starts from the same profile for all values of $x$. We see the formation of
the dense layer for small $x$, and the evolution towards a smoother
profile for large $x$.
}
\label{fig:proftpscourts}
\end{figure}

\begin{figure}[h]
\centerline{
       \psfig{figure=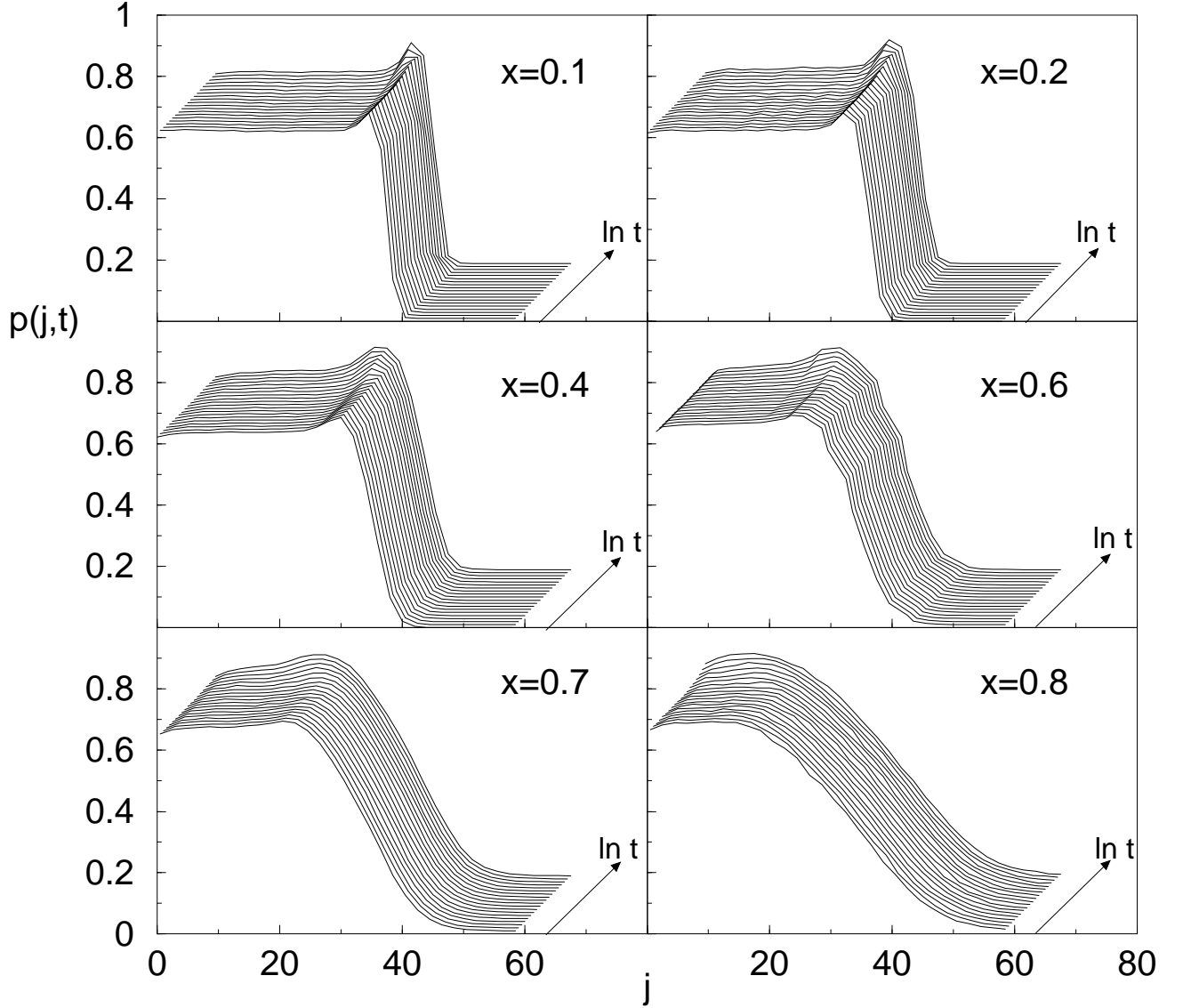,width=15cm,angle=-90} 
  \vspace{0.5cm}}
\caption{Density profiles for various values of $x$, evolving
in time ($t=0$ to $t=10^6$) after a waiting time $t_w=10^4$. The interface
is broader for higher $x$, and the dense layer below the interface
is more pronounced and evolving more slowly for lower $x$. Comparing
with Fig.~\ref{fig:proftpscourts}, we see that the overall
evolution is much slower in all cases.
}
\label{fig:prof}
\end{figure}

We have monitored the density profiles $p(j,t)$, as defined in 
eq.~\ref{profile}, of the system and of its replica after $t_w$.
While the curves of the evolution of the bulk density 
(or of other global quantities like $C$ or $B$) have
similar shapes for all values of $x$ (Eq.~\ref{eq:rho} and
Fig.~\ref{fig:rho}), the density profiles can exhibit
very different behaviours. 
In Fig.~\ref{fig:proftpscourts}, we display the short-time
($t=0$ to $t=10^4$)
evolution of the profile for several values of the shaking amplitude,
starting from the same profile for all values of $x$.
In Fig.~\ref{fig:prof}, we display the successive evolution,
for $t_w < t < t_w+10^6$ with $t_w=10^4$.
The first observation is that the width of the interface is
larger for higher shaking amplitudes, which is a quite intuitive result.
Moreover, we see that, for small shaking amplitudes, a very dense layer
forms just under the interface. This layer (whose height is of $5$ 
to $10$ lattice sites) is able to block the compaction process: 
in order to make the system more compact particles have to
rearrange also in the bulk, and the dense layer acts as a barrier. The
bottom part of the sample is therefore almost not evolving, as shown
in Fig.~\ref{fig:prof}.
As time evolves the dense layer becomes broader, 
though in a very slow way.
In all cases, the comparison of Fig.~\ref{fig:proftpscourts} and
\ref{fig:prof} also shows that the short-time dynamics is much faster
than the successive evolution.

We note that these results are in agreement with experimental results
showing that, at shaking amplitude not too large, the compaction
is more efficient in the higher parts of the media (see
first reference in \cite{exp-compaction}) and that the locally measured 
density is larger in higher parts of the sample.

For higher $x$, diffusion is easier, and the layer forming
at the top is less dense, broader, and evolves
faster (see in Fig.~\ref{fig:prof} the evolution of the dense layer
for $x=0.1,\ 0.2,\ 0.4,\ 0.6$); the compaction process
in the bulk is therefore facilitated: thus, even if the interface
is broader, which could indicate a less dense system, the bulk will
be able to compactify much better. The global density can therefore
be in fact enhanced.
This explains that, at least at finite times, the curves of
the density can be higher for higher $x$, while one would expect 
the contrary starting from a dense system; of course, for very 
strong shaking,  the system is very loose and the interface is 
very wide: this explains the existence of an
optimal shaking amplitude for compaction (see Fig.~\ref{fig:rho})
and, as mentioned before, why this optimal amplitude 
depends on the way of defining the bulk 
(it roughly corresponds to the value of $x$ for which the 
interface attains the defined bulk). 

We have again an indication of the importance of looking at the 
density profile: two very different profiles can have the same bulk 
density or potential energy (mean height).

Let us now turn to the analysis of the response function.
As described in the introduction, we let the system evolve at constant
$x$ during a certain waiting time $t_w$, and then make a copy of
the system, which we submit to a slightly different shaking,
i.e. $x+dx$. We have mostly used $dx=0.01$, and checked with 
$dx=0.005$ and $dx=0.02$ that the system was in the linear response regime
(especially for $x=0.1$ for which $dx=0.01$ is $10 \%$ of $x$).

The response function has been introduced in Sect. II through 
eq.~\ref{eq:rho}: it is defined as the difference between the potential 
energy of the two copies of the system. 
Therefore, a {\it positive} response means that the perturbed
system (at higher $x$) has a higher energy, and therefore it is 
{\it less compact} than the unperturbed one. This is what one 
clearly expect at high $x$ for example, where the system is very 
loose and a higher $x$ means that the particles are less subject to gravity.
On the other hand, a {\it negative} response means that the perturbed
system is {\it more compact} than the unperturbed one. This is not to
exclude a priori, since we have already seen in Fig.~\ref{fig:rho}
that the density at fixed time is not a monotonous function
of $x$.

\begin{figure}[h]
\centerline{
       \epsfig{figure=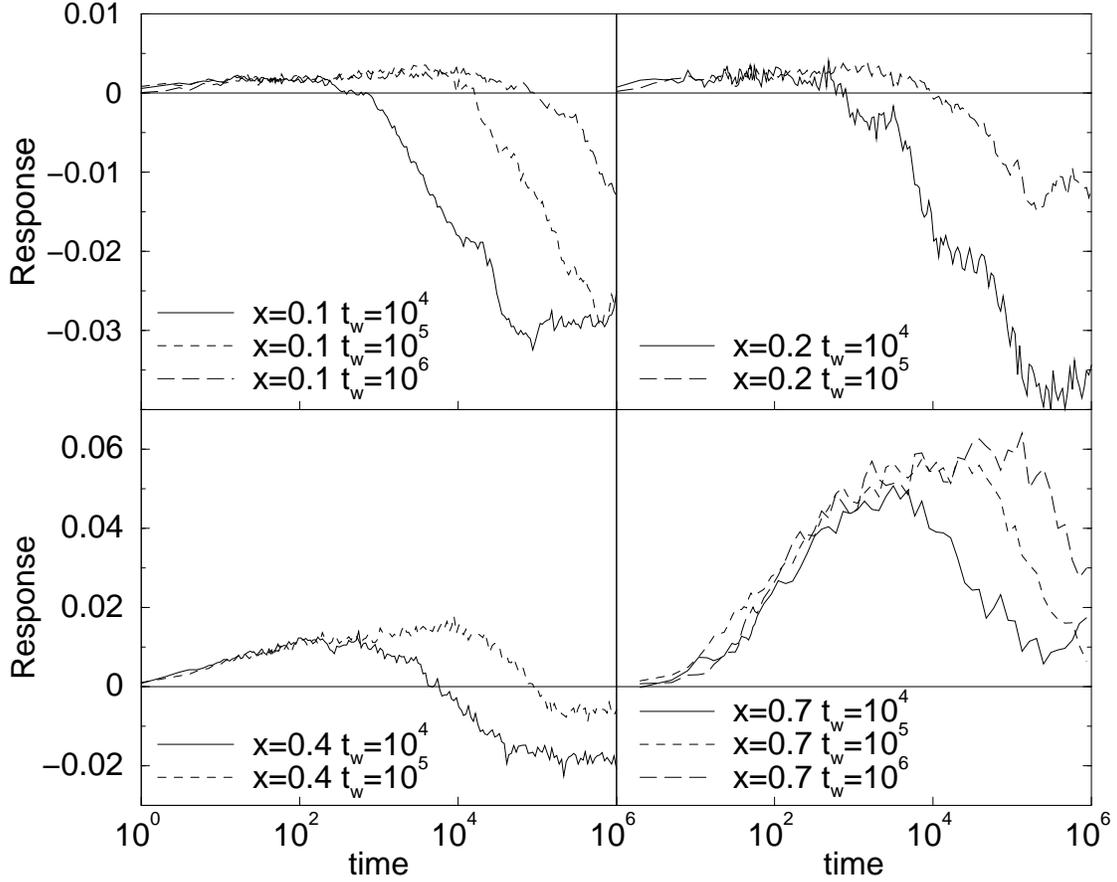,width=12cm,angle=-90}
             \vspace{0.5cm}}
\caption{Response function $R(t+t_w,t_w)$ versus $t$, for several
values of $x$ and $t_w$. The vertical scale is the same for the
left and right parts. The response tends to remains positive for a longer 
time higher are either $x$ or $t_w$.}
\label{fig:rep}
\end{figure}

For $x=0.1$, we see indeed (Fig.~\ref{fig:rep})
that the response exhibits first a positive branch followed, after a 
certain time that depends on $x$ and $t_w$, by a negative branch.
The system is first relatively decompactified by the perturbation 
(positive response), but then, at later times, it compactifies better. 
This phenomenon was first noted in the case of the Tetris
model in \cite{nicodemi}, where however only small values of $t_w$ 
were used, and more emphasis was given to the negative part. 
For a fixed $x$, however, as $t_w$ grows, the positive part extends 
to longer times and cannot be neglected.

\begin{figure}[h]
\centerline{
       \epsfig{figure=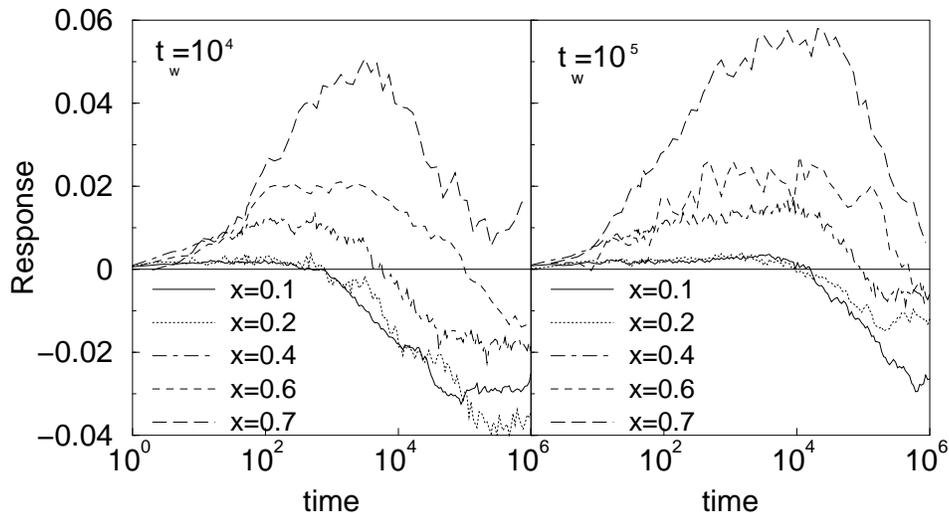,width=7cm,angle=-90}
  \vspace{0.5cm} }
\caption{Response function $R(t+t_w,t_w)$ versus $t$, for
$t_w=10^4$ (left) and $t_w=10^5$ (right), showing that, at fixed $t_w$,
the response is an increasing function of $x$.}
\label{fig:rep2}
\end{figure}

As $x$ grows, the positive part gets a higher amplitude and extends
to longer times, and for $x \ge 0.7$
no negative response can be reached (Fig.~\ref{fig:rep} and \ref{fig:rep2}).

This apparently odd behaviour can be understood by looking at the 
density profiles, and especially at the differences of the profiles between 
the perturbed and unperturbed systems, i.e. at the spatial distribution 
of the response. The response is made by two main contributions.
On the one hand the interface gives a positive response, i.e. a larger $x$ will 
naturally lead to a looser interface, because a larger $x$
means higher probability for the particle to move upwards. 
On the other hand, however, the effect on the bulk is less obvious, 
since the particles can be blocked by the ones situated above 
(as is the case at small $x$, with the dense layer appearing),
and need a global rearrangement of other particles in order to be able 
to move.

\begin{figure}[h]
\centerline{
       \psfig{figure=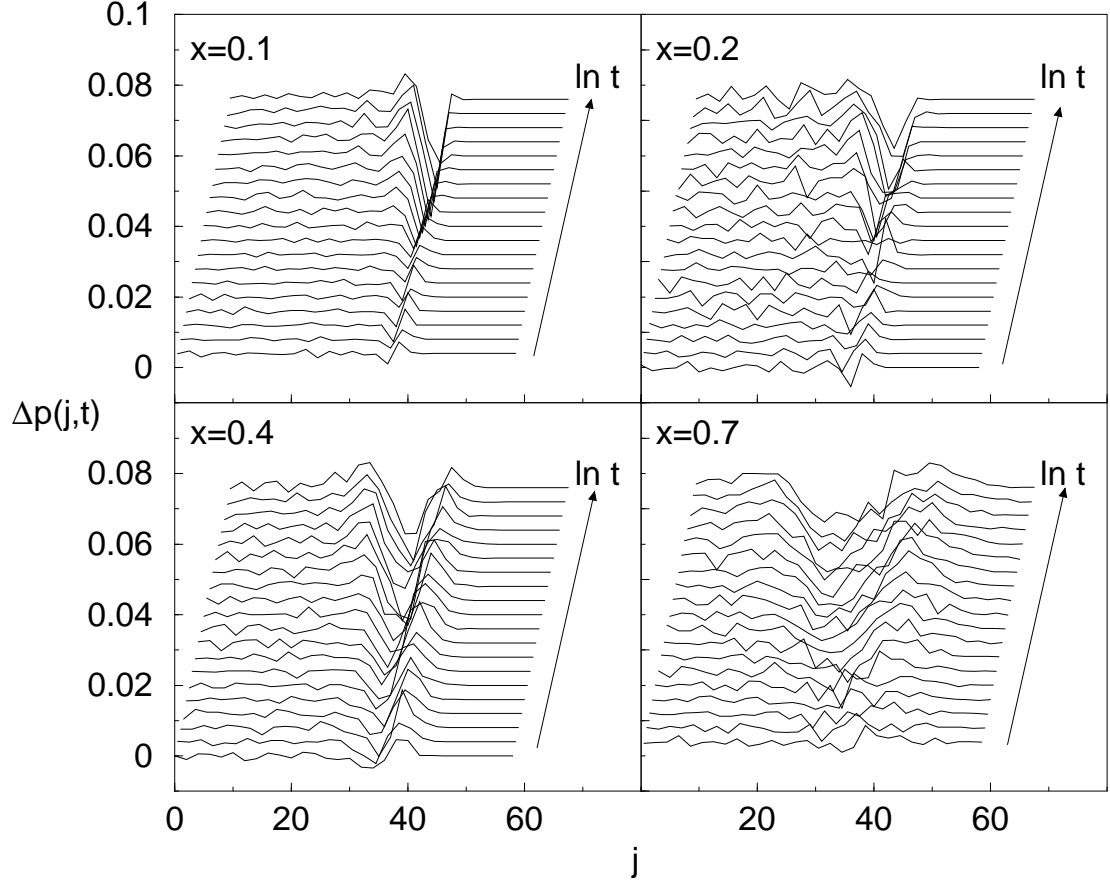,width=12cm,angle=-90}
  \vspace{0.5cm}}
\caption{Temporal evolution ($t=0$ to $t=10^6$) of the difference between 
the density  profiles of the perturbed and unperturbed systems, 
$\Delta p(j,t+t_w)$, after $t_w=10^4$ and various values of $x$. 
Comparing this figure with Fig.~\ref{fig:prof} one observes that the dips 
and bumps are located near the interface, and they are larger for larger 
$x$, i.e. for a broader interface.}
\label{fig:deltaprof}
\end{figure}

We show in Fig.~\ref{fig:deltaprof} how the difference
$\Delta p(j,t+t_w)=p^r(j,t+t_w) - p(j,t+t_w)$ 
as a function of $j$ evolves in time after $t_w$. 
A positive $\Delta p(j,t+t_w)$ means that the perturbation has locally
(at height $j$) compactified better (giving a local
positive contribution to the response function), 
while a negative $\Delta p(j,t+t_w)$ corresponds to a locally 
less compact perturbed system (local negative contribution to the 
response function). Besides, recalling the definition of the response 
function, $R(t+t_w,t_w)=\sum_j \Delta p(j,t+t_w) (j+1) L/N_{part}$, 
it is important to remark that the values of $\Delta p(j,t+t_w)$ 
coming from the interface (large $j$) bring a strong contribution to the 
response due to the term $(j+1)$.

\begin{figure}[h]
\centerline{
       \psfig{figure=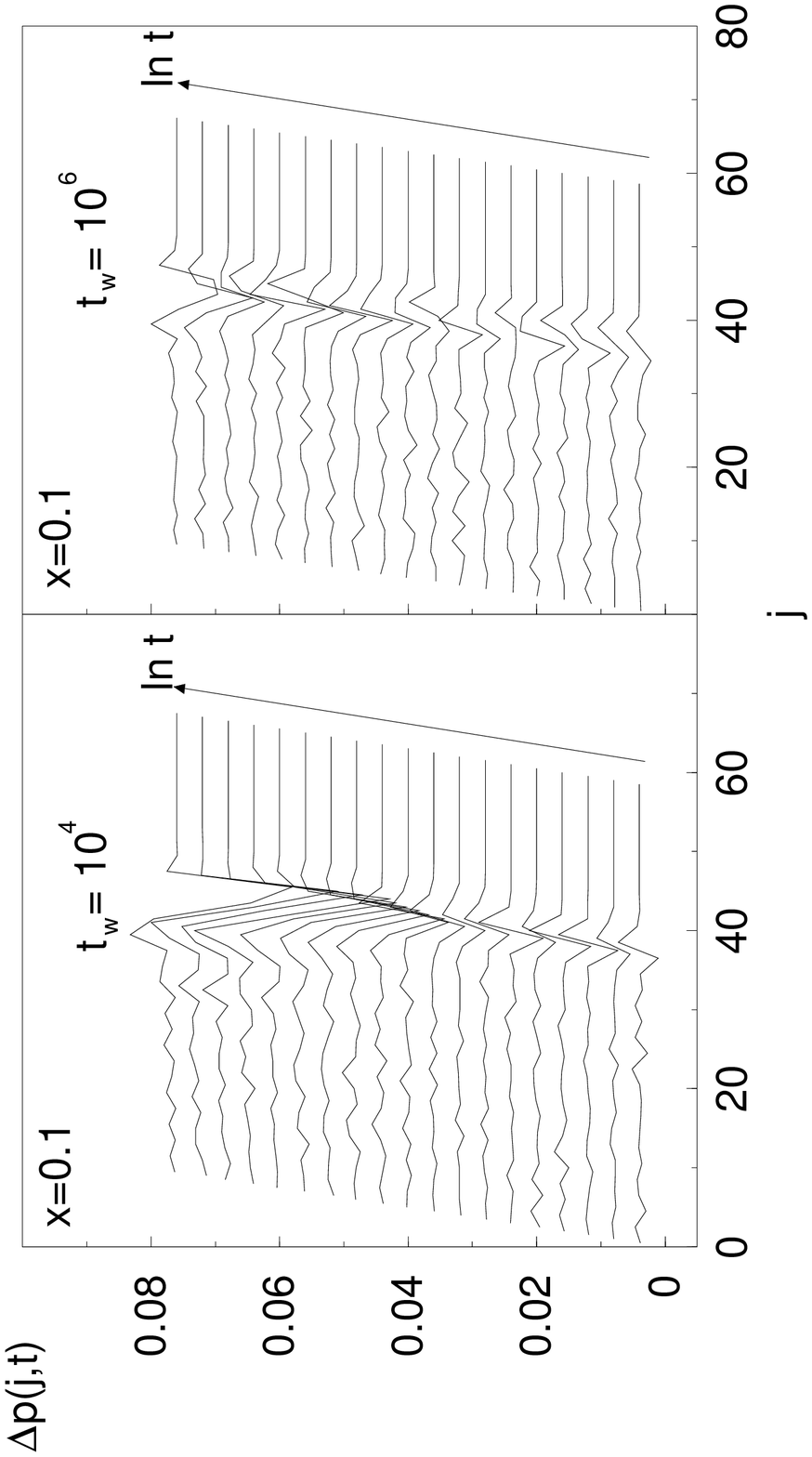,width=7cm,angle=-90}
  \vspace{0.5cm}}
\caption{Temporal evolution ($t=t_w$ to $t=t_w+10^6$) of the difference 
between the density profiles of the perturbed and unperturbed systems, 
$\Delta p(j,t+t_w)$, for $x=0.1$ and two waiting times: $t_w=10^4$ and 
$t_w=10^6$. The larger is $t_w$, the slower is the overall processes.}
\label{fig:deltaprof2}
\end{figure}

At low $x$ ($x=0.1$), the bulk is blocked by the thick dense layer just
below the interface, as shown in Fig.~\ref{fig:prof}.
Therefore, the first effect
of the perturbation $dx$ is just to decompactify the interface, which
gives rise to a positive $R$. However, once the interface is loosened,
particles in the bulk may be allowed to rearrange more freely,
and we obtain therefore
a better compaction of the bulk, and a negative $R$. 
This phenomenon is shown in Fig.~\ref{fig:deltaprof}
by the creation of a dip at heights $j$ just below the interface,
with a bump $p^{r}(j,t) > p(j,t)$ just above the interface at 
low $t$, and then by the decrease of this bump with a mass transfer 
toward the low-$j$ values of the dip. 
Let us remark, however, that a negative response,
which means a better compaction for the perturbed replica, 
does not give any substantial change in the profile: 
the response to the perturbation is very heterogeneous, 
and the bottom part of the sample simply does not feel it.
As $t_w$ grows, the interface and the dense layer become more and more
compact, and therefore harder and harder to decompactify. This explains
why the response function stays positive for longer times: the
decompaction process takes longer and longer (Fig.~\ref{fig:deltaprof2}).

At higher values of $x$, the interface becomes smoother, the blocking
layer becomes less dense and the effect of the perturbation on the 
interface becomes stronger and stronger.
At strong shaking, i.e. large values of $x$, only the interface 
contributes to the response function, which therefore is positive at 
all times. We see in Fig.~\ref{fig:deltaprof} that the 
effect of the perturbation is to transfer particles upwards.

In summary, as either $t_w$ or $x$ grows, the bulk is more
compact, and therefore its contribution to the response is smaller. 
The response function tends to become positive due to the contribution 
of the interface.

\section{Response properties for a cyclic shaking procedure}

In this section we consider a shaking procedure defined as a sequence
of steps where $x$ is varied following a cycle or a more general 
function.
A generic procedure is defined giving an initial and a final value
of $x$, $x_I$ and $x_F$, 
the maximal value of $x$, $x_{max}$, the value
of the increments in $x$ (in our simulations we have 
always used $\Delta x =0.01$) and the 
time interval $\Delta \tau$ the system spends at each value of $x$.
We have considered values of $\Delta \tau$ equal 
to $10^2,\ 10^3,\ 10^4$ and $10^5$.
The two main procedures we have considered  are defined in the following way:
\begin{itemize}
\item {\em cycle}: increase from $x_I=0.01$ 
to $x_{max}=0.8$, and then decrease to $x_F$.

\item {\em cooling}: simple decrease from 
$x_I = x_{max}$ ($x_I=0.6,\ 0.7,\ 0.8,\ 0.9$) to $x_F$.
\end{itemize}
The final values of $x$ were $x_F=0.4,\ 0.2,\ 0.1$.
Once the final value $x_F$ of $x$ was reached, $x$ is kept constant
and the measurements
of the response function (with a copy evolving at $x+dx$), 
the correlation function and the density profiles were
done either without any further delay, or with an additional
waiting time at constant $x=x_F$ of $10^4$ or $10^5$ time-steps.

We shall see that such procedures, similar to a slow cooling for thermal 
systems, allow to reach large densities, seemingly unreachable
at constant $x$. We note that these procedures are indeed the ones
used experimentally to efficiently compactify granular systems.
The examination of the density profiles will allow to gain a deeper 
insight for the effectiveness of such cycles.

\begin{figure}[h]
\centerline{
       \psfig{figure=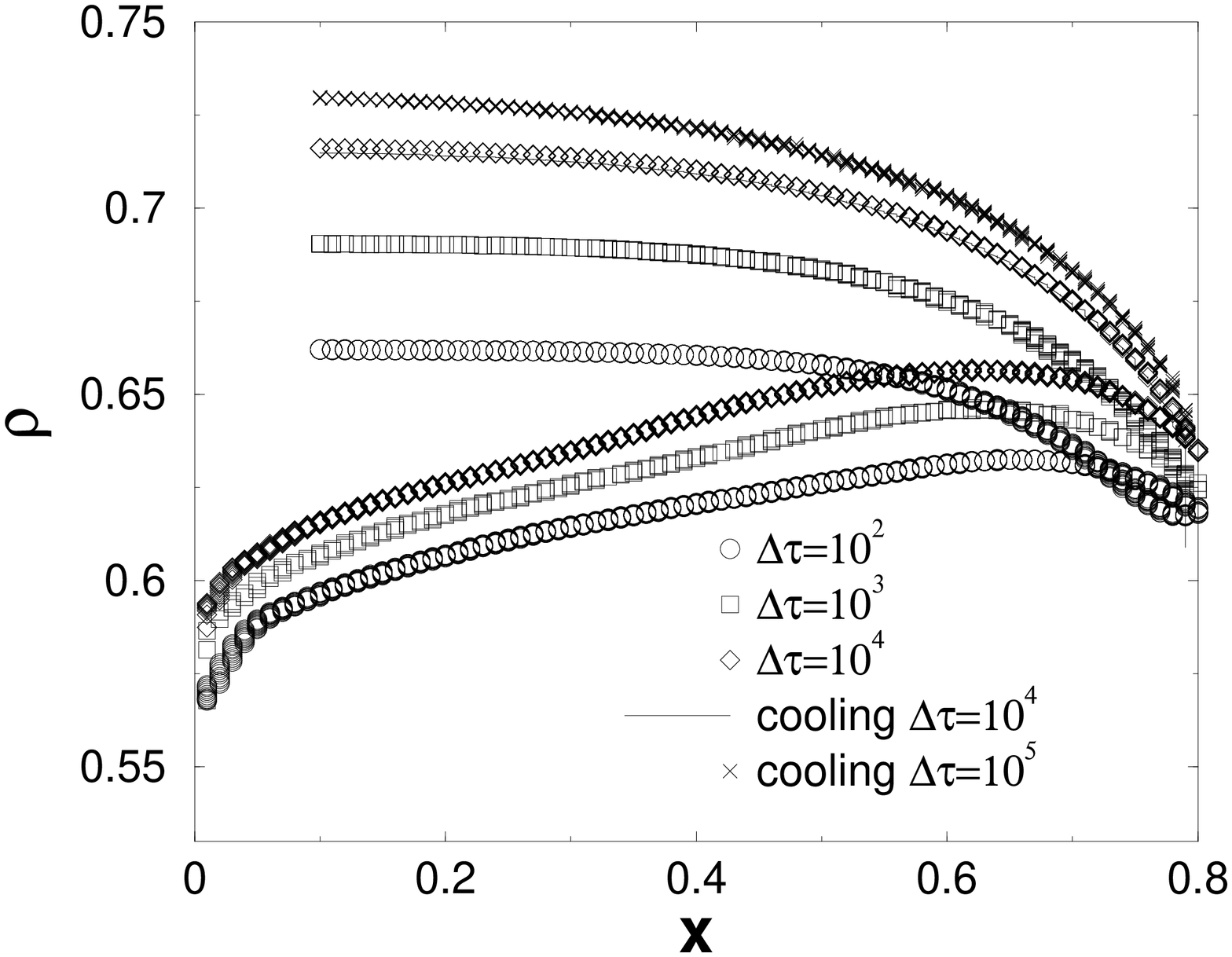,width=6cm,angle=-90}
       \hspace{1.0cm}
       \psfig{figure=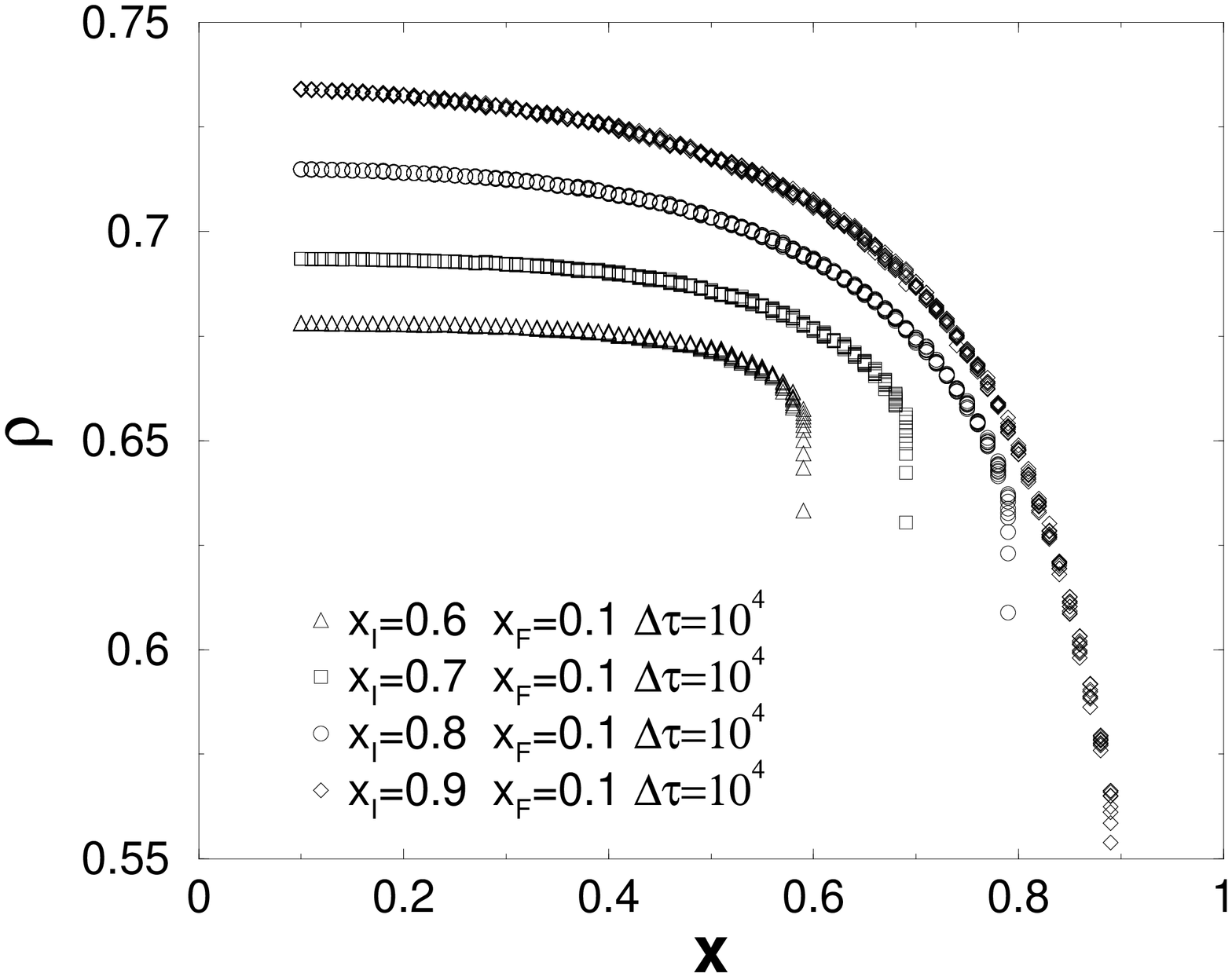,width=6cm,angle=-90}
  \vspace{0.5cm}}
\caption{Bulk density versus $x$ during the cycle (left)  
or cooling (right) procedures, for various values of $\Delta \tau$ 
and $x_{max}$. On the left figure is shown for comparison the 
behaviour of two cooling procedures with $\Delta \tau= 10^4$ and $10^5$.
We see that a simple cooling from $x_{max}$ is equivalent to a cycle 
with the same $x_{max}$. In general higher densities are obtained with 
slower procedures (larger $\Delta \tau$) or higher values of 
$x_{max} = x_I$.}
\label{fig:rhocycle}
\end{figure}

We first plot in Fig.~\ref{fig:rhocycle} the evolution of the density 
during a cycle; the influence of $\Delta \tau$ is clear, and similar 
to the case of cooling: larger $\Delta \tau$ allow to reach higher densities.
Moreover, the obtained densities are impressively higher than the densities
obtained at constant shaking amplitude.

We also checked that the part of the cycle with growing $x$ has no
practical use or influence on the ``cooling'' part 
(see Fig.~\ref{fig:rhocycle}). 
Only the maximal value of $x$ is relevant. This seems reasonable, since
structures formed at low $x$ are destroyed by a shaking at larger $x$.
On the other hand, if $\Delta \tau$ is not very large, we have
observed that a second cycle can be useful to obtain still higher 
densities. The right part of Fig.~\ref{fig:rhocycle} also shows 
that higher densities are obtained with higher values of $x_{max}$.

\begin{figure}[h]
\centerline{
       \psfig{figure=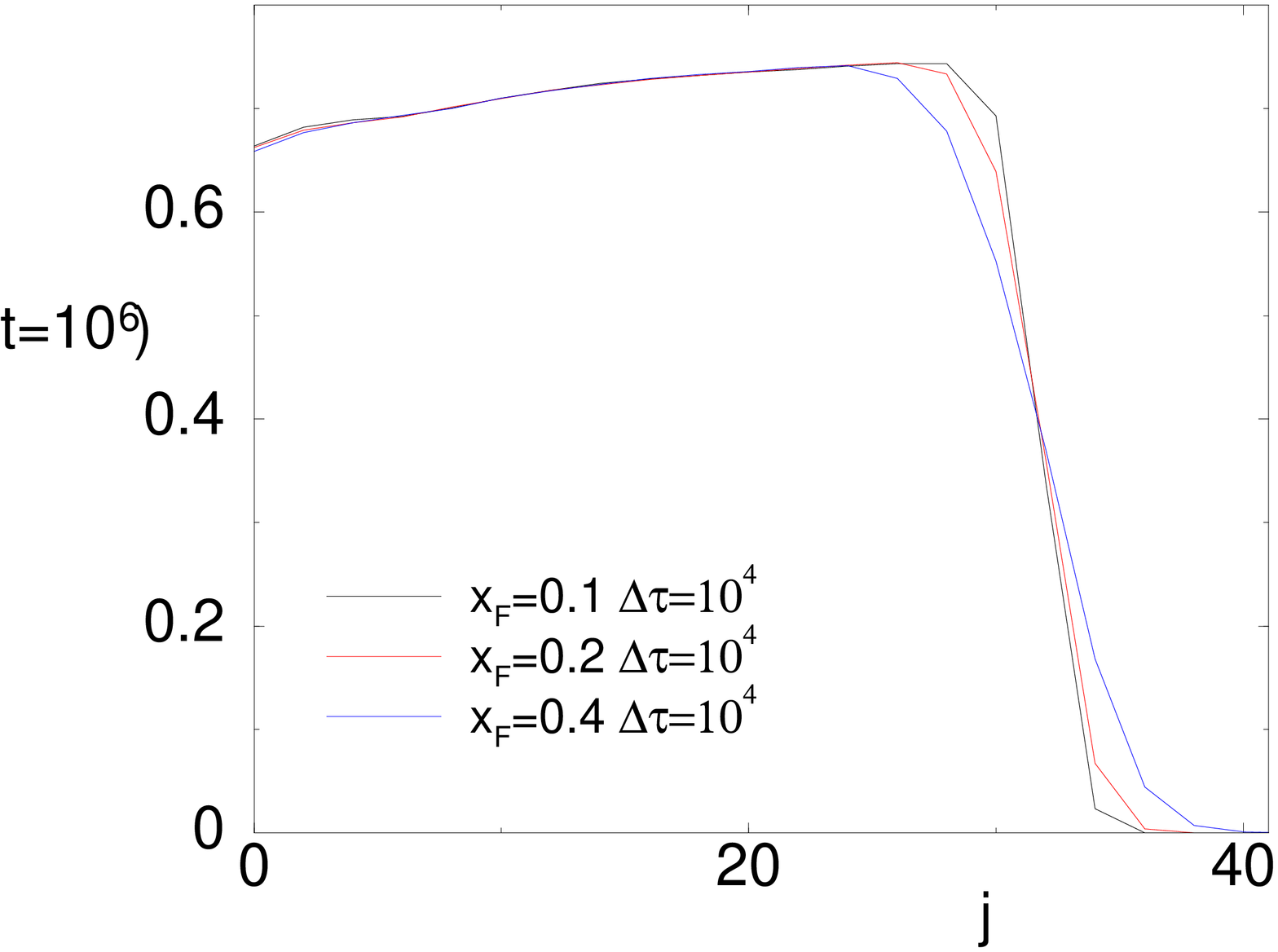,width=6cm,angle=-90}
       \psfig{figure=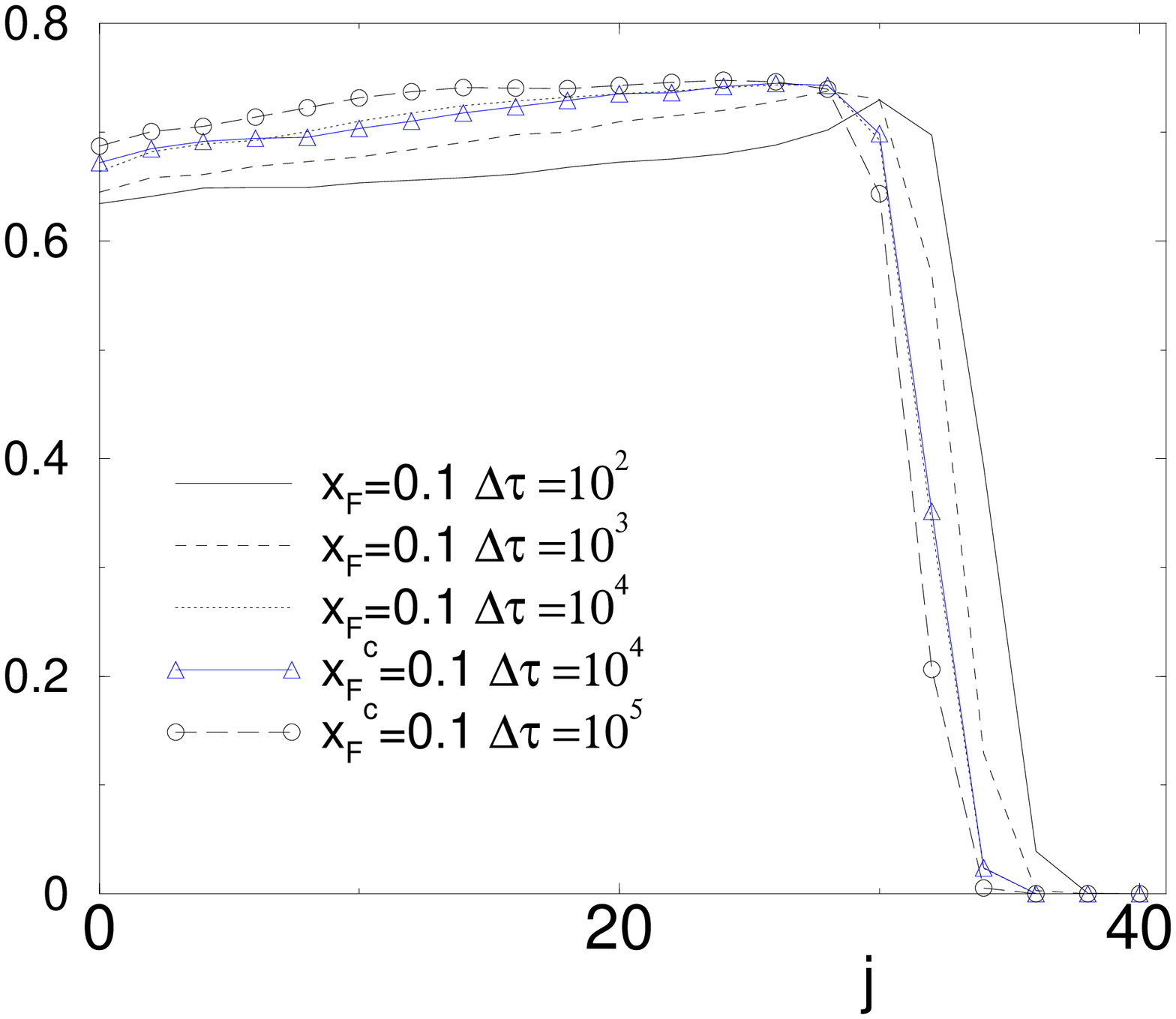,width=6cm,angle=-90}
  \vspace{0.5cm}}
\centerline{
       \psfig{figure=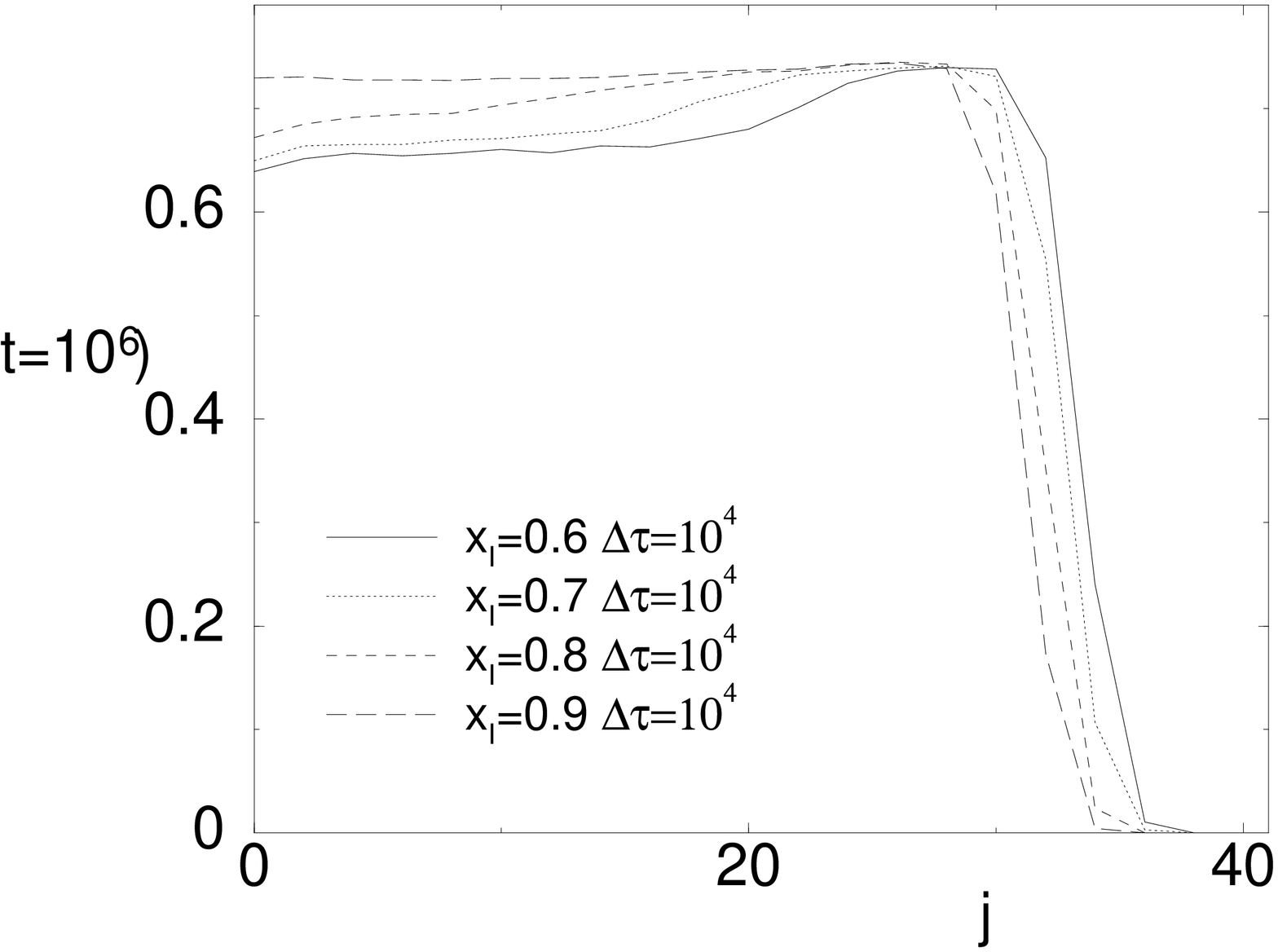,width=6cm,angle=-90}
       \psfig{figure=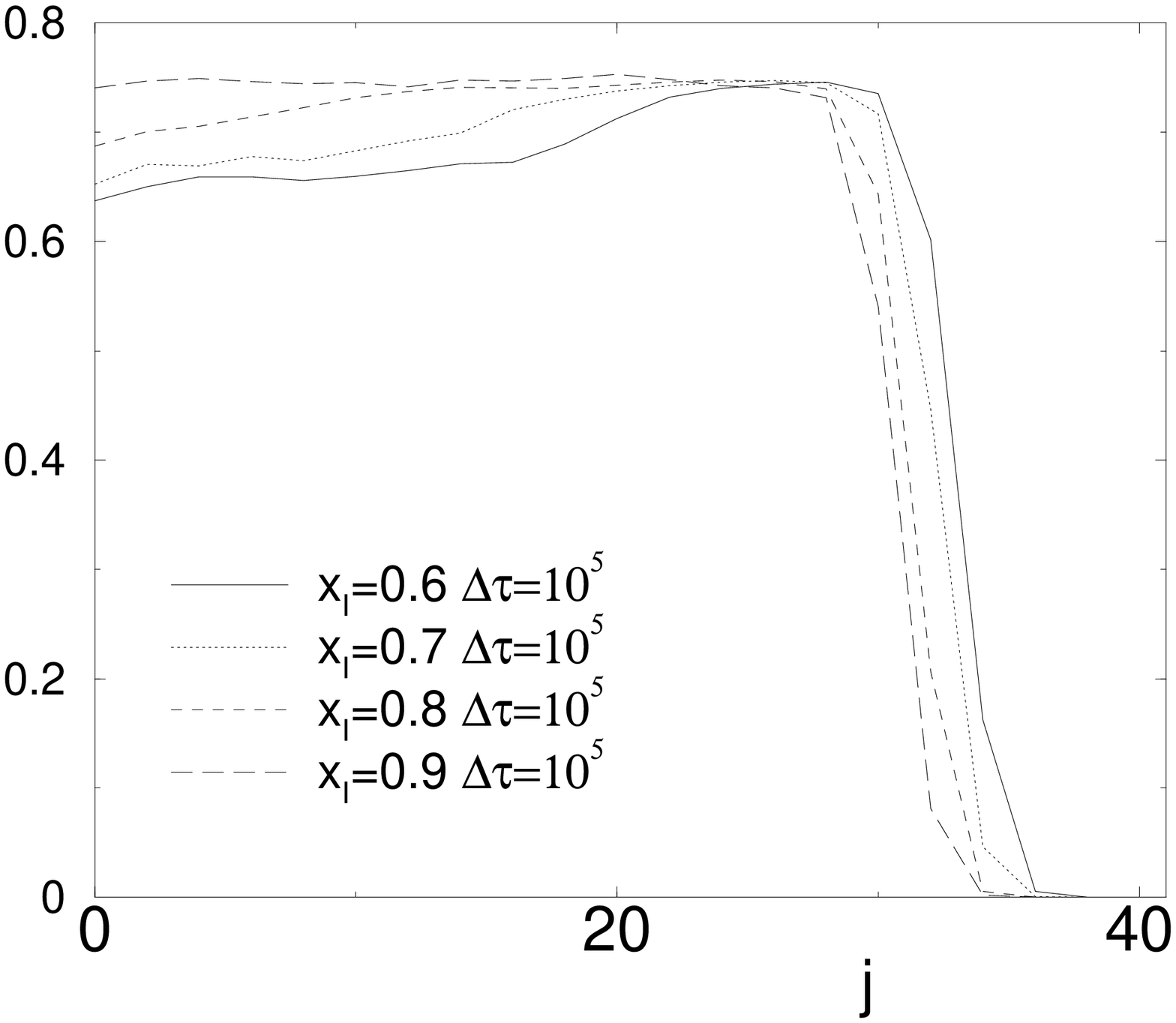,width=6cm,angle=-90}
  \vspace{0.5cm}}
\caption{
{\em Top left}: 
influence of $x_F$ on the density profile after a cycle, for 
$x_{max}=0.8$, $\Delta \tau = 10^4$; the bulk is not affected, 
but the interface is steeper for smaller $x_F$. 
{\em Top right}: 
at constant $x_{max}=0.8$ and $x_F=0.1$, effect of $\Delta \tau$; 
the bulk is denser for larger $\Delta \tau$, while the steepness 
of the interface is not changed; $x_F=0.1$ corresponds to a complete 
cycle, while $x_F^c=0.1$ corresponds to a cooling from $x_{max}=0.8$. 
We see that the two procedures are equivalent.
{\em Bottom left}: 
influence of $x_I=x_{max}$, at constant $\Delta \tau=10^4$; the bulk
is denser for higher $x_I$, the steepness of the interface is not changed.
{\em Bottom right}: 
same for $\Delta \tau=10^5$.}
\label{fig:prof2}
\end{figure}

It is quite interesting to look at the density profiles at 
various $x_F$ after a cycle (see Fig.~\ref{fig:prof2}).
For $x_F=0.1$, the bulk is much denser than after
any reachable $t_w$ at constant $x$, and the interface is as steep.
(Note however that, for $\Delta \tau$ too small, there is still a dense
layer at the interface for $x_F=0.1$.)
Moreover, at fixed $x_I$, a larger $\Delta \tau$ yields better compaction
in the bulk, as does, at fixed $\Delta \tau$, a larger $x_I$.

The comparison of the profiles at $x_F=0.4,\ 0.2$ and $0.1$ (at fixed
$x_I$ and $\Delta \tau$) shows that the bulk parts of the profiles are 
identical: only the interfaces change and they are steeper for lower $x$.
When $x$ is lowered, the bulk retains its properties while the interface
is gradually sharpened. This means that,
in order to better compactify, one has to take into account that
high values of $x$ are effective for the bulk while low values
of $x$ make the interface denser and steeper.
 Besides, the larger $x_I$ (see bottom pictures of Fig.~\ref{fig:prof2}), 
the deeper the bulk is affected at the beginning of the cooling, and the 
more compact the system is at the end of the cooling.

From these observations we deduce that the optimal method to compactify is
to begin with a high $x_I$, and decrease $x$ as slowly as possible.
If $x_I$ is higher, then one is allowed to take a lower
value of $\Delta \tau$, thus gaining time in the compaction process.
Note moreover that, for $x_I=0.9$, the difference between
the profiles obtained at $\Delta \tau=10^4$ or $10^5$ is very small.
Moreover, if only the bulk has to be compactified, one can stop the process
at intermediate values of $x$ (see Fig.~\ref{fig:prof2}) 
while, in order to also have a sharp interface,
the process has to be continued down to low values of $x$.
It would be interesting to have detailed experimental tests of these 
predictions.

\begin{figure}[h]
\centerline{
       \psfig{figure=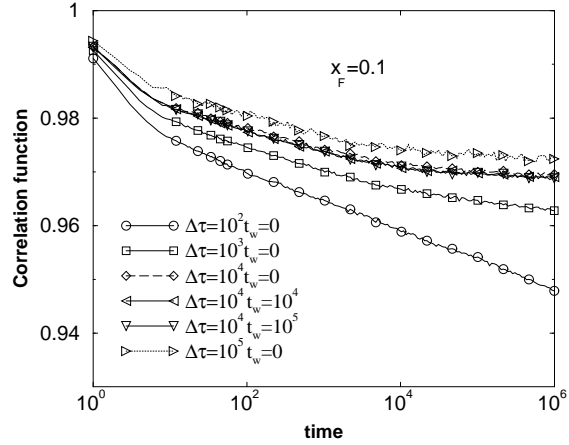,width=6cm,angle=-90} 
  \vspace{0.5cm}}
\caption{Correlation function after a cycle with $x_F=0.1$, $x_{max}=0.8$.
The longer the cycle, the older the system looks (i.e. the slower
the correlation decays). The curves for $\Delta \tau=10^4$ with
$t_w=0,\  10^4,\ 10^5$ are perfectly collapsed, showing 
time-translation invariance.}
\label{fig:corr_cycle}
\end{figure}

After a cycle or a cooling procedure, the mass-mass two-times 
correlation function (Fig.~\ref{fig:corr_cycle}) 
consistently shows an effective age of the system much larger 
than  the real one: for example, for a global real time of 
$7 \cdot 10^5$ (cooling from $x_I=0.8$ to $x_I=0.1$ in steps of
$\Delta x=0.01$ with $\Delta \tau=10^4$), the effective age 
is much larger than $10^6$ (largest waiting time simulated at
constant $x$).
This fact is confirmed if we wait an additional
$t_w=10^4$ or $10^5$ after the cycle: we observe time-translation invariance,
showing that the system is in an equilibrium or quasi-equilibrium state.

\begin{figure}[h]
\centerline{
       \psfig{figure=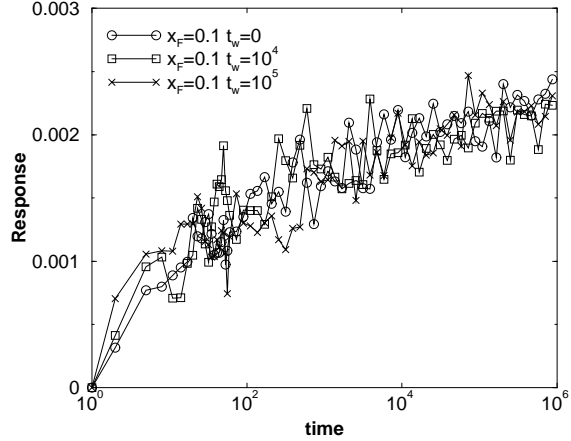,width=6cm,angle=-90} 
  \vspace{0.5cm}}
\caption{Response function after a cycle with $x_F=0.1$ and $x_F=0.8$,
$x_{max}=0.8$, and a time $t_w=0,\ 10^4,\ 10^5$ at constant
$x=x_F=0.1$. The response is small with large fluctuations, but
definitely positive. We show only the case $x_F=0.1$, the response being
an increasing function of $x_F$.}
\label{fig:rep_cycle}
\end{figure}

The response function, shown in Fig.~\ref{fig:rep_cycle},
is also independent of $t_w$ (for $t_w=0,\ 10^4,\ 10^5$), and
is always positive. 

\begin{figure}[h]
\centerline{
\psfig{figure=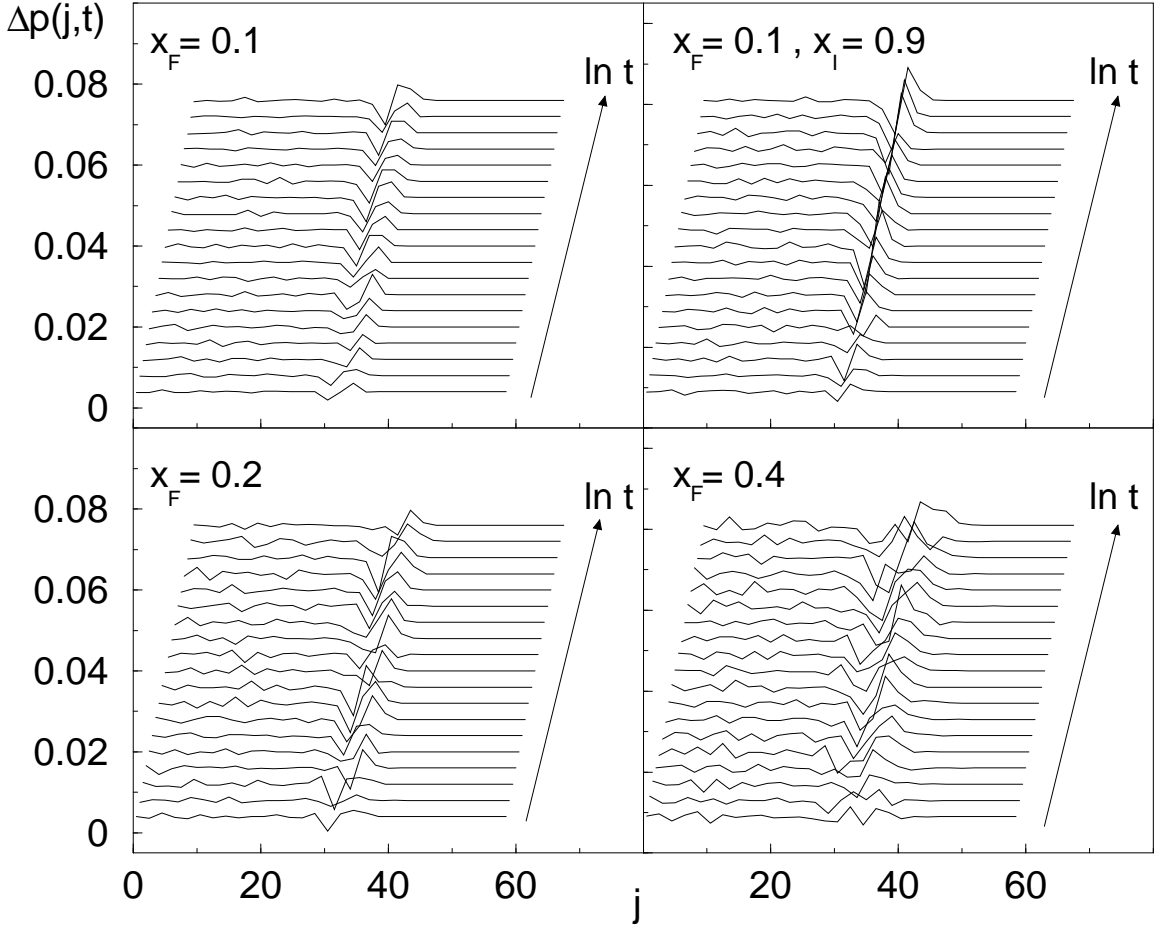,width=12cm,angle=-90}
  \vspace{0.5cm}}
\caption{Temporal evolution (from $t=0$ to $t=10^6$) of the 
difference between the density profiles of the perturbed and 
unperturbed systems, $\Delta p(j,t)$, after a cooling from 
$x_I=0.8$ to $x_F=0.1,\ 0.2,\ 0.4$, with $\Delta \tau=10^4$, 
(top left, bottom left and bottom right respectively) or from 
$x_I=0.9$ to $x_F=0.1$ (with $\Delta \tau=10^4$) (top right).}
\label{fig:profdifcycle}
\end{figure}

Once again, this behaviour can be understood by the analysis of the 
difference of the density profiles of the two replicas.
Fig.~\ref{fig:profdifcycle}  shows
$\Delta p(j,t)$ for several combinations of $x_I$ and $x_F$.
This figure is to be compared with Fig.~ \ref{fig:deltaprof}.
In all the cases one observes that the change in $x_F$ only modifies the
interface, broadening it.  The interface in this case is always composed by
a denser layer (bump) lying above a looser layer (dip).
The only effect of the perturbation is thus a broadening of the interface, 
even at small $x_F$, thus causing a positive response, because the
bulk has been already efficiently compactified by the cycling
procedure.

\section{Comparing different procedures: the importance of the history}

In this section we try to give a unitary view for the results we have 
shown so far. Since in a granular system there is no dynamics without 
any energy injection, i.e. without any external perturbations, it is 
highly important to deeply understand the response properties to such 
perturbations. One of the most important point to stress is the fact 
that the response is never homogeneous. One can never assume that the 
properties of the system are homogeneously distributed, not even assuming 
a coarse grained point of view. The perturbations drive
the system into an instability mechanism that generates large-scale spatial 
structures \cite{baldassarri,valanghe}. Let us discuss what are the 
consequences of this phenomenon that has been named {\em Self-Organized 
Structuring}\cite{valanghe}. 

First of all, one is not allowed to describe a static packing in terms only of 
a scalar quantity like the density. It is evident how it is possible 
to construct several different packings corresponding to the same 
global average density, each one with completely different rheological 
properties (for a concrete example concerning the behaviour of a granular 
system subject to shearing see \cite{dilatancy}). Recent experiments
\cite{joss} have made this point clear. Depending on the 
properties investigated one is then forced to enlarge the parameter 
space in order to give a reasonable description of the system. One 
important ingredient to take into account is the history, e.g. the 
ensemble of dynamical procedures the system has undergone till the 
moment where we analyze it. One can then ask where the information 
about the past history is encoded and whether is possible to take 
into account this history in some suitable coarsed grained view. 
One possibility is to consider the density as a local parameter 
and describe the system in terms of some density map. It is precisely 
in this spirit that in this paper we have considered, as a first step 
in this direction, the analysis of the time evolution of the density 
profiles \cite{dens_map}. This corresponds to investigating the 
properties of the system in the direction of the imposed external 
field: the gravity.

The presence of large-scale structures in the system makes
the problem of the response properties far from being trivial.
One of the first question one could ask concerns the best procedure
to compactify a given sample. The best strategy refers to the strategy 
that allows to obtain the highest density in a given time.
Alternatively one could consider the best strategy to reach a given density
in the smallest time.  Since different parts of the sample
(in the zero-th order schematization one can consider the bulk and the 
surface) respond in a different way to different values of the shaking 
amplitude $x$, it is obvious to conclude that the best strategy will
not coincide with a procedure in which one keeps $x$ fixed indefinitely.
In this case, in fact, as we have shown in Sect. III, the average 
bulk density at fixed time is not a monotonous function of $x$.
It is then quite intuitive to figure out different procedures
where $x$ is a complex function of time. We have shown in Sect. IV that
one of the best procedures is to consider a cooling process where one 
starts with a relatively high value of $x=x_{max}$ and decrease 
progressively $x$ with a rate $1/{\Delta \tau}$ up to a final value
$x_F$. The higher are either $x_{max}$ or ${\Delta \tau}$, the larger
will be the final asymptotic density. The smaller is the value of
$x_F$, the larger will be the compactified region and the sharper 
the interface.

The rationale behind this definition of the optimal compaction
procedure can be understood as follows. The best way to compactify
globally the system is to start from the compaction of the bottom part 
of the system. In order to do this one should choose high values of 
$x$ which allow to extend the interface of the system, i.e. the mobilized 
region, deeper and deeper. The procedure then proceeds reducing 
progressively the value of $x$ in order to compactify sequentially
regions at larger heights. It is clear that a small value of $x$
cannot affect the part of the system already compactified with a 
larger value of $x$. That is why in order to better compactify also 
the system interface one has to continue the procedure to very small
values of $x$. In this way one can associate to each particular value
of $x$ during the procedure the optimal compaction of a specific 
region of the system.
On the other hand it is clear why the compaction process gives better
results for larger $\Delta \tau$. Spending longer on a certain value of
$x$ allows to better compactify the region corresponding to this value
of $x$.
The shaking procedure at constant $x$ cannot thus be effective because
a large $x$ will only be able to compactify the deep bulk while a small
$x$ will create a high density layer below the interface (see 
Fig.~\ref{fig:prof}) which will stop for a very long time the bulk compaction.

The presence of large-scale structures draws a lot of consequences
also in the behaviour of the response function as defined in 
eq.~\ref{response}.  From the results presented in the previous sections, 
it is evident as, even in a very simplified picture, the sign of the 
response  function depends on a complex convolution of several contributions: 
the spatial structures spontaneously emerging in the system (again at the 
zero-th order of approximation  one can consider the sum of two coupled 
contributions coming from the surface and the bulk), the value of the 
shaking amplitude and the past history of the system (encoded in the value 
of $t_w$ for a constant shaking amplitude or, more generally, in the 
whole definition of the dynamical procedure). It is then evident how 
trying to explain the results about the sign of the response function 
on the only basis of the shaking amplitude ($x$) can be fallacious. 
There does not exist, as suggested in \cite{nicodemi}, a transition in 
$x$ such that the the response function is positive above a certain value 
of $x$ and negative below this value.
In the case of constant shaking amplitude, no matter what the value of 
$x$ is, one typically observes a positive response due to the contribution 
of the system surface, eventually followed by a negative regime at times 
that depend on $x$ and $t_w$: the larger are $x$ or $t_w$, the
larger is the time 
over which one sees a positive response. This is due to the complex 
balance between the contributions coming from the surface and the ones 
from the bulk. These features were not considered in \cite{nicodemi} 
where the role of inhomogeneities was neglected and where only 
small values of $t_w$ were considered. 
On the other hand if one looks at the response 
after a cyclic procedure, as described in sect. IV, one realizes that 
the response is always positive because in this case only the
interface is giving an important contribution.

It is interesting to compare the results obtained for the mass-mass
two-times correlation function using different driving procedures. 
We have shown in Sect. III the presence 
of ageing in a system driven with a given constant $x$. In this case 
one observes a two-steps relaxation of the correlation function, typically
observed in glassy systems. The second relaxation is quite well
described by a function decaying as $log(t_w)/log(t+t_w)$ no matter the
value of $x$. Similar behaviour has been observed for the mean-square
distance between the potential energies, $B(t+t_w,t_w)$.
On the other hand, the behaviour of the correlation function
turns out to be completely different if the system is subject
to a more complicate driving procedure. In Sect. IV we have shown 
that the use of a cyclic or of cooling procedure brings the system in a 
state where time-translation invariance holds, i.e. there is not anymore 
ageing. Experiments analyzing the presence or absence of
ageing in granular material would certainly be welcome, and a first
step is being taken in \cite{joss}.
In this case the state reached by the system is such that
the bulk is almost completely decoupled from the system interface,
i.e. the bulk density does not change anymore with time. At this stage
the dynamics of the system is concentrated on the interface which is almost
in an equilibrium state compatible with the final value of $x$, $x_F$.
The system does not age anymore and the global response 
(see Fig.~\ref{fig:rep_cycle}) is always positive. 
For another approach to the relaxation dynamics of granular media
that focuses on the asymptotic stationary state reached 
after a very long shaking procedure we refer the reader to
\cite{caglioti_loreto}.

These last observations arise a further question concerning
the possibility to associate to a granular system a unique scalar 
temperature when the space translation invariance is broken.
In \cite{caglioti_loreto} it has been shown that this is indeed 
possible if the system is in a stationary state. In this case 
one can show that it there exists a ``temperature''-like quantity
which is made uniform everywhere in the system and which is related 
do the derivative of a suitable free-energy-like functional.
It is evident how one gets immediately into troubles if the 
system is far from being homogeneous and if the presence of
spatial structures is accompanied by a breakdown of the 
time-translation invariance (ageing behaviour).
The above discussion about the sign of the response function 
represents an example of how data analysis can be misleading.
It is then clear how in general one cannot describe the response 
properties of a granular system in terms of a unique parameter, 
e.g.  the temperature. Considering different values of 
$x$ or $t_w$ or considering different dynamical procedures, 
one is simply exploring different non-equilibrium regions of 
the phase-space that correspond to different spatial inhomogeneities. 
From the above discussion it should be clear that a negative value 
of the response function does not mean that one can associate
a negative effective temperature to the whole system\cite{nicodemi}
(in the spirit of the Fluctuation-Dissipation Theorem (FDT)
\cite{eff_temp}). In this case one has simply that for the specific 
values of $x$ and $t_w$ explored, the contribution coming from the bulk
is important.
This contribution can change depending on the global history
(which is encoded in the spatial structures), on $x$ and $t_w$ and 
thus, unless one is able to introduce more subtle, and local, indicators, 
it seems to us quite meaningless and misleading to try to extend
to this case the definition of an effective temperature.
From this point of view, unless one is able to define suitable 
free-energy-like functionals\cite{edwards} taking into account 
the heterogeneities of the system, any statement about the validity of the 
Fluctuation-Dissipation Theorem (FDT) in such an heterogeneous 
systems \cite{nicodemi}, statement which anyway has to be considered 
in the limit $t_w \rightarrow \infty$, seems to us hazardous.

\section{Conclusions} 

In this paper we have investigated the response properties 
of granular media in the framework of a recently proposed class 
of models, the so-called Random Tetris Model.
On the one hand, we focused our attention on global quantities as
the global density, the response and the correlation functions.
On the other hand we have monitored some local quantities that allowed
us to investigate the large-scale structures spontaneously emerging
in these systems as a response to the imposed perturbation
(driving). The comparison between global and local quantities allowed
us to gain a deeper insight on how granular materials respond to 
perturbations and in this perspective of the importance 
of spatial structures.  We have considered several different perturbation
procedures defined in terms of the temporal functions describing the
shaking amplitude ($\Gamma$ in the experiments and $x$ with 
$\Gamma \simeq 1 / log(1/\sqrt x)$ in the models).
In particular we have analyzed the case where one keeps
$x$ indefinitely constant and compared this case to the one 
where $x$ varies as a function of time, $x(t)$.
Our main results can be summarized as follows. In the case of a 
procedure at constant $x$ the system exhibits ageing described by a 
correlation function decaying as $log(t_w)/log(t+t_w)$. The response
function exhibits a complex behaviour depending on $x$ and $t_w$.
In general one always observes a positive response eventually
followed, at times increasing with either $x$ or $t_w$, by a negative
response. All these properties can be explained looking at the 
heterogeneities arising in the density  profiles. 

The scenario changes completely considering more complex shaking 
procedures where $x=x(t)$.
In this case (see Sect. IV for the details) the system can be found
in an almost stationary state where ageing is not anymore present, i.e.
the correlation function is time-translation invariant, and the 
response function is always positive. Also in this case the comparison
of the results with the analysis of the density profiles allows
us to gain a deeper understanding of the effect of the perturbation
on the system. On this basis we are able to formulate some specific
recipes for the best compaction procedure and to comment on some recent
results concerning the validity of the Fluctuation-Dissipation Theorem 
and the possibility of a thermodynamic description for these
non-thermal systems.

Let us conclude with two points that open possibilities for future
work: (i) recent experiments \cite{joss} have focused on the response,
not to a slight change in the forcing, but rather to a large
change, in the spirit of experiments of temperature cycling in
spin-glasses \cite{tcycling}. Work is in progress to reproduce
the preliminary experimental results and draw comparisons
with the spin-glasses phenomenology.
(ii) we have defined the response function, as in \cite{nicodemi},
as the response of the system to a change in the driving force, 
which is itself considered as an analog of the temperature. 
In order to try to understand
if an extension of the FDT can be thought of, it would probably be more
reasonable to look at the response to a force acting randomly on the
particles, and in a way uncorrelated to the overall driving, in a way
similar to the one used in a lattice-gas model \cite{sellitto}
or in models of super-cooled liquids \cite{parisi-barrat-kob}.

Needless to say finally
that it would be extremely important to have an experimental
check of our predictions to be used as a starting point for further
theoretical investigations.

{Acknowledgments:} We thank C. Josserand and P. Viot for discussions
and communication of results prior to publication.
VL acknowledges financial support under 
project ERBFMBICT961220. This work has been partially supported 
from the European Network-Fractals under contract No. FMRXCT980183.


\begin{thebibliography}{90}

\bibitem[*]{umr}Unit{\'e} Mixte de Recherche UMR 8627.

\bibitem{grain} For a recent introduction to the overall phenomenology
see Proceedings of the NATO Advanced Study Institute on
 {\it Physics of Dry Granular Media}, Eds.
 H. J. Herrmann {\it et al}, Kluwer Academic Publishers, Netherlands
 (1998). 

\bibitem{RTM}  E. Caglioti, S. Krishnamurthy and V. Loreto,
{\em Random Tetris Model}, unpublished (1999). 

\bibitem{prltetris}
E. Caglioti, V. Loreto, H.J. Herrmann and M. Nicodemi,
{\em Phys. Rev. Lett.} {\bf 79}, 1575 (1997).

\bibitem{segtet}
E. Caglioti, A. Coniglio, H.J. Herrmann, V. Loreto and M. Nicodemi,
{\em Europhys. Lett.} {\bf 43}, 591 (1998).

\bibitem{dilatancy} M. Piccioni, V. Loreto and S. Roux,
in press in {\em Phys. Rev. E} (1999).

\bibitem{nicodemi_coniglio} M. Nicodemi, A. Coniglio, 
{\em Phys. Rev. Lett.} {\bf 82}, 916 (1999).


\bibitem{valanghe} S. Krishnamurthy, H. J. Herrmann, V. Loreto,
M. Nicodemi and S. Roux, {\em Fractals} {\bf Vol. 7} No.1 (1999) 51-58;

S. Krishnamurthy, V. Loreto, H.J. Herrmann, S.S. Manna and
S. Roux, {\em Phys. Rev. Lett.} {\bf 83}, 304 (1999);

S. Krishnamurthy, V. Loreto and S. Roux,
{\em Bubbling and Large-Scale Structures in Avalanche Dynamics},
subm. to {\em Phys. Rev. Lett.} (1999). 

\bibitem{baldassarri} A. Baldassarri, S. Krishnamurthy, H.J. Herrmann, 
V. Loreto and S. Roux, 
{\em Coarsening and Slow-relaxation in Granular Compaction}, 
in preparation (1999).

\bibitem{nicodemi} M. Nicodemi, {\em Phys. Rev. Lett.} 
{\bf 82}, 3734 (1999).

\bibitem{exp-compaction}
Knight, J. B., Fandrich, C. G., Lau, C. N., Jaeger, H. M. 
and Nagel, S. R. {\em Phys. Rev. E} {\bf 51}, 3957 (1995);

Nowak, E. R., Knight, J. B., Povinelli, M., Jaeger, H. M. 
and Nagel, S. R., {\em Powder Technol.} {\bf 94}, 79-83 (1997);

Nowak, E. R., Knight, J. B., Ben-Naim, E., Jaeger, H. M. 
and Nagel, S. R., {\em Phys. Rev. E} {\bf 57}, 1971-1982 (1998);

Jaeger, H. M. in Physics of Dry Granular Media (eds. Herrmann, H.
J., Hovi, J.-P. and Luding, S.) 553-583 (Kluwer Academic, Dordrecht, The
Netherlands, 1998).

\bibitem{viot}
J. Talbot, G. Tarjus and P. Viot, preprint cond-mat/9910239.

\bibitem{exp2d}
J. Duran, T. Mazozi, E. Cl\'ement and J. Rajchenbach,
{\em Phys. Rev. E} {\bf 50}, 3092 (1994).

\bibitem{mod-compaction} see for instance: 

T. Boutreux and P.G. de Gennes, in
{\em Powders and Grains 97}, edited by R. Behringer and J. Jenkins
(A.A. Balkema, Rotterdam), pp.439.;
 
E. Ben-Naim, J.B. Knight and E.R. Nowak,
{\em Physica D} {\bf 123}, 380 (1998); 

P.L Krapivsky and E. Ben-Naim,
{\em J. of Chem. Phys.} {\bf 100}, 6778 (1994);
 
M. Nicodemi, A. Coniglio, H.J. Herrmann,
{\em Phys. Rev. E}, {\bf 55}, 1 (1997).


\bibitem{nota2} $\rho_\infty$ depends on $x$ and is not $1$ as 
in the simplest Tetris model \cite{prltetris} where the existence 
of a ground state with a perfect antiferromagnetic ordering allows
the configuration at density $1$. 
In the RTM the random choice of the particles removes 
this limitation.

\bibitem{review_aging}
See for example
J.-P. Bouchaud, L.F. Cugliandolo, J. Kurchan and M. M\'ezard,
in {\em Spin Glasses and Random Field} ed. A. P. Young,
(World Scientific, 1997).

\bibitem{bouchaud} J.-P. Bouchaud, {\em J. Physique I} 
(France) {\bf 2}, 265 (1995).

\bibitem{viot_private} P. Viot: priv. comm. (1999).

\bibitem{joss} C. Josserand, in preparation (1999).

\bibitem{dens_map} A further step in this direction is to consider the real 
two-dimensional density maps. For this we refer the reader to 
\cite{RTM,baldassarri}.


\bibitem{caglioti_loreto} E. Caglioti and V. Loreto, 
{\em Phys. Rev. Lett.} {\bf xx}, xxx, (1999).

\bibitem{eff_temp} L.~F. Cugliandolo, J.~Kurchan and L.~Peliti
{\em Phys. Rev. E} {\bf 55}, 3898 (1997).

\bibitem{edwards} It is interesting in this perspective to look at: 
S.F.~Edwards, in {\em Current Trends in the Physics of Materials}, 
(Italian Physical Society and North Holland, Amsterdam, 1990);
 S.F.~Edwards  and R.B.S.~Oakeshott,
{\em Physica A} {\bf 157}, 1080 (1989);
 S.F.~Edwards and C.C.~Mounfield,
    {\em Physica A} {\bf 210}, 279 (1994).      

\bibitem{tcycling} See e.g. E. Vincent,
J. Hammann, M. Ocio, J.-P. Bouchaud, L.F. Cugliandolo,
in {\em Spin Glasses and Random Field} ed. A. P. Young,
(World Scientific, 1997)  and references therein.

\bibitem{sellitto} M. Sellitto, {\em Eur. Phys. J. B} {\bf 4}, 135 (1998).

\bibitem{parisi-barrat-kob} See e.g.
G. Parisi, {\em Phys. Rev. Lett.} {\bf 79} 3660 (1997); 
W. Kob and J.-L. Barrat, {\em Eur. Phys. J. B} (1999) in press, 
{\em Physica A} {\bf 263}, 234 (1999). 

\end{thebibliography}
\end{document}